\documentclass[default,iicol]{sn-jnl}

\usepackage{amsmath}
\usepackage{stmaryrd} 
\usepackage{enumitem} 

\usepackage{array}
\usepackage{multirow}

\jyear{2021}%

\theoremstyle{thmstyleone}%
\theoremstyle{thmstyletwo}%

\theoremstyle{thmstylethree}%

\usepackage{color} 
\newcommand*{\revision }{\color{black} }

\begin{document}

\title[Quantitative phase microscopies]{Quantitative phase microscopies: accuracy comparison}


\email{guillaume.baffou@fresnel.fr}
\author[1]{\fnm{Patrick C.} \sur{Chaumet}}
\author[2]{\fnm{Pierre} \sur{Bon}}
\author[1]{\fnm{Guillaume} \sur{Maire}}
\author[1]{\fnm{Anne} \sur{Sentenac}}
\author*[1,3]{\fnm{Guillaume} \sur{Baffou}}

\affil[1]{Aix Marseille Univ, CNRS, Centrale Marseille, Institut Fresnel, Marseille, France}
\affil[2]{Universit\'e de Limoges, CNRS, XLIM, UMR 7252, F-87000 Limoges, France}
\affil[3]{Neurotechnology Center, Department of Biological Sciences, Columbia University, New York, NY 10027, USA}


\abstract{{\revision Quantitative phase microscopies (QPMs) play a pivotal role in bio-imaging, offering unique insights that complement fluorescence imaging. They provide essential data on mass distribution and transport, inaccessible to fluorescence techniques. Additionally, QPMs are label-free, eliminating concerns of photobleaching and phototoxicity. However, navigating through the array of available QPM techniques can be complex, making it challenging to select the most suitable one for a particular application.} This article presents a thorough comparison of themain QPM techniques, focusing on their accuracy in terms of measurement precision and trueness. We focus on 8 techniques, namely digital holographic microscopy (DHM), cross-grating wavefront microscopy (CGM), which is based on QLSI (quadriwave lateral shearing interferometry), diffraction phase microscopy (DPM), differential phase-contrast (DPC) microscopy, phase-shifting interferometry (PSI) imaging, Fourier phase microscopy (FPM), spatial light interference microscopy (SLIM), and transport-of-intensity equation (TIE) imaging. For this purpose, we used a home-made numerical toolbox based on discrete dipole approximation (IF-DDA). This toolbox is designed to compute the electromagnetic field at the sample plane of a microscope,  irrespective of the object's complexity or the illumination conditions. We upgraded this toolbox to enable it to model any type of QPM, and to take into account shot noise. In a nutshell, the results show that DHM and PSI are inherently free from artefacts and rather suffer from coherent noise; In CGM, DPC, DPM and TIE, there is a trade off between precision and trueness, which can be balanced by varying one experimental parameter; FPM and SLIM suffer from inherent artefacts that cannot be discarded experimentally in most cases, making the techniques not quantitative especially for large objects covering a large part of the field of view, such as eukaryotic cells.}

\keywords{Quantitative phase microscopy, wavefront microscopy, precision, accuracy, trueness, DDA, simulations}



\maketitle

\section{Introduction}
Quantitative phase imaging (QPI) refers to a family of optical imaging techniques that aim at imaging the phase of a light beam \cite{Book_Popescu,S13_4170,NP12_578,OLE135_106188,ACSNano16_11516}, a quantity that is normally not accessible using common optical sensors. QPI has witnessed remarkable progress these last 2 decades, fueled by advances in optical instrumentation and computing technology. QPI techniques have been heavily used in optical \emph{microscopy}, leading to the closely-related field of quantitative phase microscopy (QPM), and significantly expanded the capabilities of microscopy in cell biology. One of the main strength of QPM in cell biology is its ability to measure the dry mass of cells, due to the close relation that exists between refractive index and mass density \cite{N169_366,JOSA47_545,PPT33_102096,NM11_1221}. The dry mass is defined as the mass of the cell excluding its water content, {\it i.e.}, the mass of biological material. Monitoring the dry mass of cells enables precise measurements of cellular growth rate and matter transport at the single cell level in a label-free, non-invasive manner.

The optical path difference (OPD) $\delta\ell$ map is the image that is often computed in practice, in QPM, rather than the phase image itself. It is defined by the optical path variation created by the imaged object once placed in the field of view of the microscope. It is simply related to the phase image $\varphi$ {\it via} the relation
\begin{equation}
\varphi=\frac{2\pi}{\lambda}\delta\ell .\label{phideltaell}
\end{equation}
The interest of the OPD is that, in first approximation, it is directly proportional to the dry mass density $\rho$ (pg/\textmu m$^2$) of live cells, in particular far from absorbing bands,  :
\begin{equation}
\rho=\gamma^{-1}\delta\ell, \label{massdensity}
\end{equation}
where $\gamma$ is called the specific refraction increment. For biological media, it is approximately constant, ranging from 0.18 to 0.21 \textmu m$^3$/pg. From the dry mass density can be computed the dry mass of the imaged cell by image segmentation and pixel summation.

{\revision Note that the dry mass is the most popular physical quantity that QPM can quantify in biology, but there exist other readouts of interest, as listed in Ref. \cite{BOE3_1757}, namely the optical volume, some phase/mass ratiometric quantities, sphericity/eccentricity indices, phase kurtosis and skewness. Also, more advanced tomography techniques can also measure cell volume and refractive index distribution in 3 dimensions \cite{Lee_Hugonnet_Park_2022,Picazo-Bueno_Mico_2023}.}

QPM methods encompass a diverse array of techniques, developed since the late 90s. Most of them consist of making interfere several beams to transform the phase information into intensity variations, measurable using common optical sensors. {\revision Albeit numerous, common QPM techniques can be classified into two main categories: \emph{off-axis} techniques and \emph{multiple-image} techniques.}

The off-axis techniques consist of making interfere two (or more) light beams impinging on a camera, tilted by a small angle to each other, to create fringes that contain the phase information. The raw image, registered by the camera, is called an interferogram, and the related techniques are off-axis DHM (digital holographic microscopy) \cite{N1_020901,AO47_A52}, DPM (diffraction phase microscopy) \cite{AOP6_57} and CGM (cross-grating wavefront microscopy) \cite{OE17_13080,ACSP10_322}.

{\revision The multiple-image techniques consist in acquiring several successive images of the object of interest under different conditions. These images are then numerically mixed to derive the wavefront or phase profile. Among them,} the phase-shifting techniques also consist of the interference between two beams, an object beam and a reference beam, but the two beams are collinear~\cite{AO27_5082}. Thus, no interference pattern (no fringes) appears on the camera. When no tilt is applied between the two beams, no bijective relation exists between the grayscale intensity on each camera pixel, and the light phase. Intensity and phase can be retrieved separately by the acquisition of 4 images associated with 4 phase shifts of 0, $\pi/2$, $\pi$ and $3\pi/2$ applied to one of the two beams. The related techniques are PSI (phase-shifting interferometry imaging) \cite{AO27_5082,AOP7_1,JM179_11,EPJAP44_29}, FPM (Fourier phase microscopy) \cite{OL29_2503,AO46_1836} and SLIM (spatial light interference microscopy) \cite{OE19_1016}. There also exist non-interferometric techniques, where multiple bright-field images are acquired, e.g., under various conditions.  The related techniques are differential phase contrast (DPC) microscopy \cite{OE23_11394,BOE7_3940,OL37_4062}, which acquires 4 images of an object obtained with 4 different illumination angles, and combine the 4 images to retrieve the phase of the sample; and TIE (transport-of-intensity equation) microscopy, also called phase-diversity, which combines 3 different images at different focuses to reconstruct the wavefront profile \cite{OC199_65}.

Selecting the right QPM for a specific application is complex, as they all serve the same purpose. Review articles have been published, but they mainly aim at reviewing the applications \cite{Book_Popescu,S13_4170,NP12_578,OLE135_106188}, not really comparing the techniques with each other. Some articles and reviews recently aimed at comparing QPM techniques \cite{OE26_17498,BOE10_2768,ACSNano16_11516}. However, they concern only few QPM techniques and are usually restricted to some particular types of objects.

\begin{figure*}
	\centering
		\includegraphics[scale=0.9]{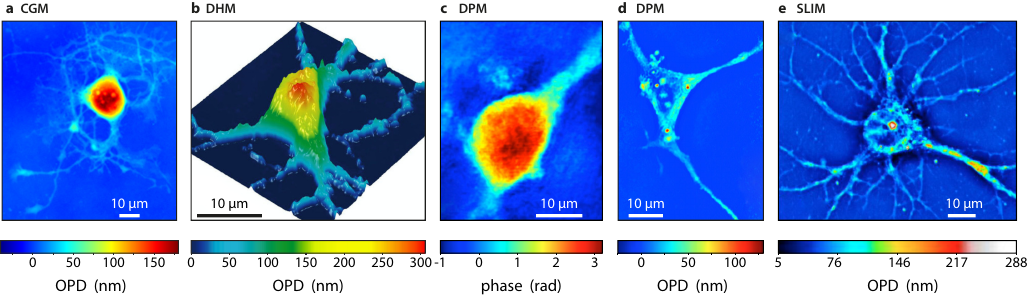}
	\caption{{\bfseries OPD and phase images of neurons acquired using different techniques} (a) OPD of a hippocampal neuron imaged using CGM/QLSI \cite{BOE13_6550}. (b) OPD of a mouse cortical neuron imaged using DHM. Reprinted with permission from \cite{N1_020901}. Copyright 2014 SPIE.  (c) Phase image of a neuron imaged using DPM. Reprinted with permission from \cite{AOP6_57}. Copyright 2014 Optical Society of America. (d) OPD image of a neuron imaged using DPM. Reprinted with permission from \cite{JB12_e201800269}. Copyright 2018 Wiley. (e) OPD of a hippocampal neuron imaged using SLIM. Reprinted with permission from \cite{OE19_1016}. Copyright 2011 Optical Society of America.}
	\label{neurons}
\end{figure*}

The other issue we recently raised \cite{ACSP10_322} is the presence of inconsistencies in the literature, where some QPMs render much different images of similar objects: Figure~\ref{neurons} displays neurons imaged using CGM, DPM and SLIM. While the images in Figs.~\ref{neurons}(a) and~\ref{neurons}(b) look consistent, featuring bumpy cell somas, the somas in images Figs.~\ref{neurons}(d) and~\ref{neurons}(e) look void. When noticing that a OPD image is supposed to represent the dry mass density, it becomes apparent that such images may not accurately reflect reality, as the cell soma contains much biological material. This raises questions about the level of accuracy achievable across QPM techniques.\\

In this article, we aim to compare the most commonly used QPM techniques, and in particular their relative degrees of accuracy, {\it i.e.} their precision and trueness. We focus on 8 QPM techniques, namely DHM, CGM/QLSI, FPM, DPM, PSI, SLIM and TIE. For the sake of comprehensiveness, we present results of numerical simulations arising from the modelling of each of the 8 microscopy setups. Our algorithm enables the computation of QPM images, including the noise amplitude and the possible presence of inherent artefacts. The first part introduces the working principles of the 8 techniques, recalls their experimental configuration, their theory and their image processing methods. The second part describes the numerical tool we developed and used to model each microscopy technique. Finally, results of numerical simulations are presented, on 4 model objects: a nanoparticle, a bacterium, a uniform slab (2D material), and a eukaryotic cell. The precision and trueness of each microscopy techniques are discussed and the origins of noise and inaccuracies are explained.

\section{QPM techniques}
This section describes the 8 QPM techniques investigated in this article, one by one, namely DHM, CGM/QLSI, DPM, DPC, FPM, PSI, SLIM and TIE. All these techniques involve a microscope, and a camera at the image plane. To describe the formalism, we shall consider monochromatic fields, using complex number notations, at the angular frequency $\omega$, defining the wavenumber $k=\omega/c$ and the wavelength $\lambda=2\pi/k$.

This section only describes the working principles of the techniques. The advantages and limitations of each technique will be discussed in the following sections.

\subsection{Bright-field microscopy}
Let us first introduce the basic experimental system and the notations with the case of the bright-field microscope (\ref{codeChart}).

We consider an object standing in the object space of the microscope. This object is illuminated in transmission by the so-called incident light beam, resulting in the generation of a scattered electric field. Let us call $E_0$ the electric field at the \emph{image} plane of a conventional microscope (no QPM for the moment) coming from the incident light, and $E_\mathrm{s}$ the electric field at the image plane scattered by the sample. The total electric field at the image plane reads thus, no matter the size and nature of the object and without approximation,
\begin{equation}
E(\mathbf{r})=E_0(\mathbf{r})+E_\mathrm{s}(\mathbf{r}).\label{EE0Es}
\end{equation}
The quantity of interest in QPM is the phase $\varphi$ of $E$, which is such that
\begin{align}
E(\mathbf{r})&=A(\mathbf{r})e^{i\varphi(\mathbf{r})},
\label{eq:EAphi}
\end{align}
where $A(\mathbf{r})=\vert E(\mathbf{r})\vert$.
\begin{figure*}
	\centering
		\includegraphics[scale=0.9]{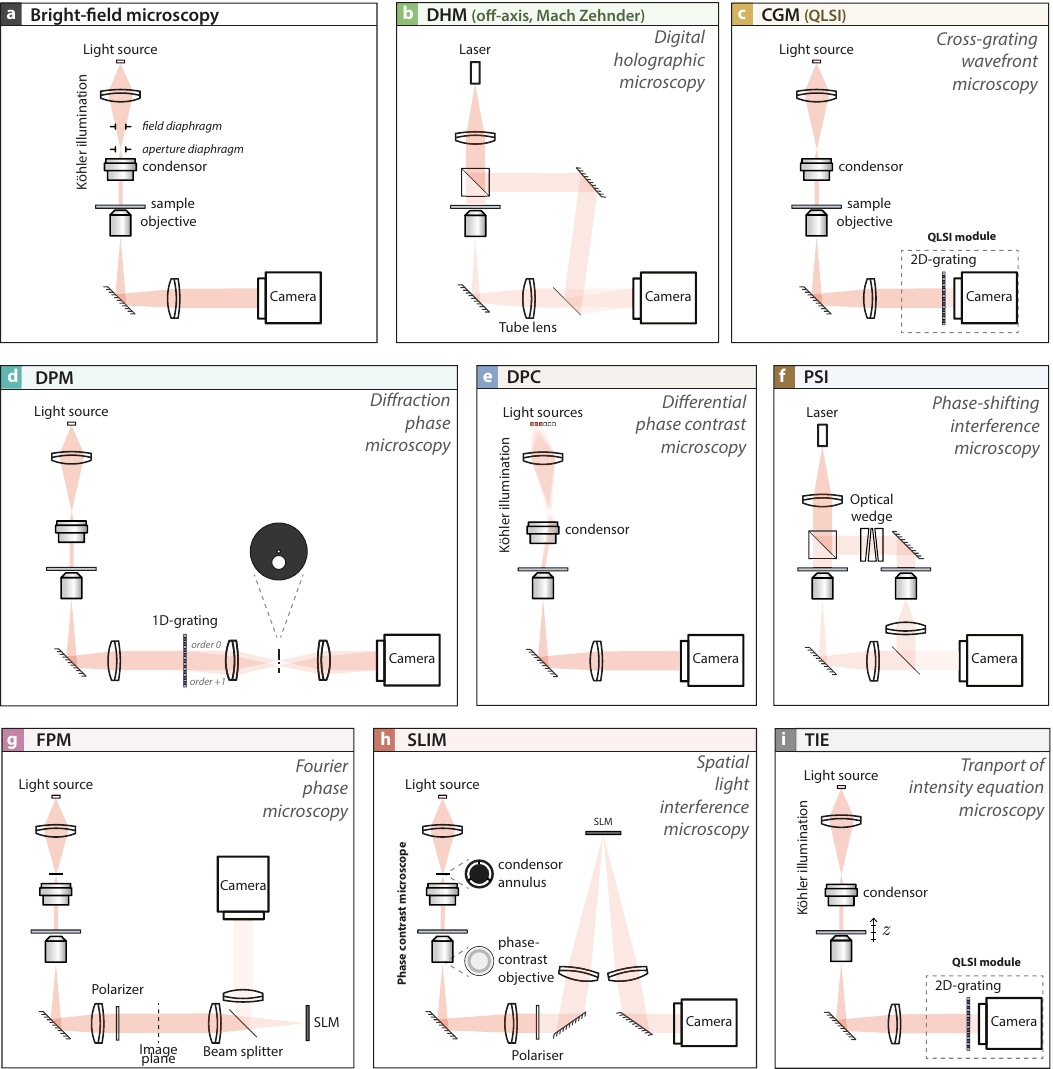}
	\caption{{\bfseries Schematics of the QPM experimental setups investigated in this study.} (a) Bright-field microscopy. (b) Digital holographic microscopy (DHM), an off-axis QPM technique, involving a reference arm. (c) Cross-grating wavefront microscopy (CGM), a technique based on the use of a 2D diffraction grating placed at a millimeter distance from a camera (an association called quadriwave lateral shearing interferometry (QLSI)). (d) Diffraction phase microscopy (DPM), an off-axis QPM based on the use of a 1D-diffraction grating placed in the image plane of the microscope, and a mask placed in the Fourier plane. (e) Differential phase-contrast (DPC) microscopy, based on the use of a series of tilted illuminations. (f) Phase-shifting interferometry (PSI), the standard phase-shifting imaging technique. (g) Fourier Phase microscopy (FPM) and (h) spatial light interference microscopy (SLIM), which are phase-shifting techniques using a spatial light modulator (SLM). (i) Transport of intensity equation (TIE) microscopy, based on the acquisition of various images acquired at various focuses. All the setups are shown in transmission because it is more common, but reflection configurations also exist \cite{Cuche_AO_99}.}
	\label{QPMsetups}
\end{figure*}

When detecting this field with a camera sensor, the measured quantity is $I=A^2$, cancelling any information on $\varphi$. The following sections describe how $\varphi$ can nevertheless be retrieved using camera sensors.\\

Note that all the equations above assume a scalar electric field. This scalar approximation in optics is usually employed and valid in two cases: (i) with a linearly polarized illumination, where the scattered light is assumed to globally retain the incident polarisation, and (ii) with an unpolarized illumination, where the two components of the electric field are likely to remain identical. However, in some cases, the scalar approximation is no longer valid, in particular when studying anisotropic particles or birefringent materials. The more general formalism involves two scalar fields, corresponding to the $x$ and $y$ components of the electric field, and Eqs. \eqref{EE0Es} and \eqref{eq:EAphi} becomes more generally:
\begin{align}
\mathbf{E}(\mathbf{r})=&\mathbf{E}_0(\mathbf{r})+\mathbf{E}_\mathrm{s}(\mathbf{r}),\label{EE0Esvect}\\
\mathbf{E}(\mathbf{r})=&A_x(\mathbf{r})e^{i\varphi_x(\mathbf{r})}\mathbf{u}_x + A_y(\mathbf{r})e^{i\varphi_y(\mathbf{r})}\mathbf{u}_y.
\label{eq:EAphivect}
\end{align}
The concept of "the phase of a light beam" is therefore ambiguous in the most general case. There are in principle 2 phase profiles. In the following sections, when describing the theory of each QPM, we will consider the scalar approximation, for the sake of simplicity.

\subsection{Off-axis techniques}
Off-axis techniques involve the interference of multiple waves, which create fringes on the camera sensor. These fringes are processed using a demodulation algorithm to retrieve the phase or wavefront of the light beam. This family includes DHM, CGM/QLSI, and DPM.

{\revision Unlike all the other techniques families described further on, off-axis techniques are based on the acquisition of a \emph{single} image, containing all the information of intensity and phase. Information theory says that if we gain information somewhere compared with bright field microscopy (here the phase), we should also loose some information elsewhere. Indeed, with this technique family, this gain in information is always accompanied by a loss in image definition (the processed images have less pixels).}

\subsubsection{Digital holographic microscopy (DHM)}
The most common QPM method is off-axis digital holographic microscopy (DHM) \cite{N1_020901,AO47_A52}. It consists of a coherent plane wave $E_\mathrm{ref}$ sent on the camera sensor at a tilted angle $\theta$ (Fig.~\ref{QPMsetups}(c)), in addition to the object beam characterized by an electric field $E$. The overlap of tilted illuminations is usually created with a Mach-Zehnder interferometry configuration. The resulting interference pattern measured by the camera, called the interferogram, consists of fringes and reads
\begin{eqnarray}
I&=&\vert E+E_\mathrm{ref}\vert^2\nonumber\\
I&=&\vert  E\vert^2+\vert E_\mathrm{ref}\vert^2+2\vert E\vert\vert E_\mathrm{ref}\vert\cos(k_xx+\varphi),\nonumber\\\label{eqDHM}
\end{eqnarray}
where $k_x=k\sin\theta$. The phase map can be extracted by a demodulation of the interferogram image using a Fourier-transform-based algorithm \cite{book_DHM_Matlab}.

\subsubsection{Cross-grating wavefront microscopy}

Cross-grating wavefront microscopy (CGM) is a microscopy technique based on the wavefront imaging technique called quadriwave lateral shearing interferometry (QLSI), invented and patented by Primot \emph{et al}. in 2000 \cite{AO39_5715,patent6577403}. CGM consists of implementing a QLSI camera in the image plane of an optical microscope (Fig.~\ref{QPMsetups}(b)) \cite{OE17_13080,ACSP10_322}. A QLSI camera is composed of a 2-dimensional diffraction grating placed at a millimeter distance from a camera sensor. A QLSI grating is designed such that it diffracts only the 1st orders, creating 4 replicas of the image, separated by a distance $a$ on the camera sensor \cite{JPDAP54_294002}. Because the 4 waves impinging on the camera are not only shifted but also tilted by the grating, by an angle $\theta$, fringes appear, just like in off-axis DHM. For the sake of simplicity and understanding in this paper, let us consider the interference pattern obtained from only 2 replicas on the camera sensor, instead of 4: \cite{ACSP10_322}:
\begin{align}
I=&\left\vert E(x+a/2)e^{ik_x x}+E(x-a/2)e^{-ik_x x}\right\vert^2\nonumber\\
I=&I_1+I_2\cos\left(\frac{4\pi}{\Gamma}x+d\nabla_x W\right),\label{eqCGM}
\end{align}
where $W$ is the wavefront profile, $k_x=k\sin\theta$, $I_1$ and $I_2$ are the intensity components, $\Gamma$ is the pitch of the grating (grexel size) and $d$ is the grating/camera distance. The 4-beam interference equation can be found in the literature \cite{OE17_13080}.

Using a very similar algorithm as off-axis DHM, not the phase information but the wavefront gradients, along $x$ and $y$, can be retrieved by two demodulations in the Fourier space \cite{JPDAP54_294002,GitHub-CGMprocess}. The wavefront profile $W$ can then be retrieved from its two gradients by numerical integration.

Note that the wavefront profile directly equals the OPD,
\begin{equation}
W=\delta\ell,\label{Wdeltaell}
\end{equation}
meaning that one does not need to convert the phase into the OPD like with other QPM techniques. In CGM/QLSI, the measured quantity is directly the quantity of interest in cell biology, proportional to the dry mass density (Eq. \eqref{massdensity}).\\

{\revision QLSI is not the only technique that is based on the positioning of an optical element at the vicinity of a camera, as a means to gain more information than $\vert\mathbf{E}\vert^2$. In particular, a microlens arrays \cite{JRS17_S573,AOP10_512} or a diffusing plate \cite{SM6_2100737} can also be placed at the vicinity of a camera. Diffuser-based phase sensing and imaging (DIPSI) uses a simple thin diffuser to retrieve the wavefront profile of a light beam \cite{OL42_5117,SM6_2100737}, with a slightly reduced image definition (i.e. pixels number) compared with QLSI. Shack-Hartmann (SH) wavefront imaging uses an array of microlenses to recover a wavefront profile \cite{JRS17_S573,OC222_81}. The image definition is even lower than QLSI or DIPSI, making SH imaging rarely used in microscopy \cite{OL42_2122}. Light field microscopy (LFM) also uses an array of microlenses, but to retrieve another hidden information: the angular distribution of the light impinging at the sensor plane \cite{AOP10_512,BCJ16_397}. We propose to refer to these techniques as \emph{sensor-proximity} techniques, {\it i.e.} techniques that place an optical element close to the camera sensor that redistributes the light intensity profile to reveal hidden information.}

\subsubsection{Diffraction phase microscopy (DPM)}
In diffraction phase microscopy (DPM), the camera is replaced by a 1-dimensional diffraction grating (Fig.~\ref{QPMsetups}(d)) \cite{AOP6_57} optically conjugated by a 4-$f$ system on the camera, placed further along the optical axis. The Fourier plane of this system can be experimentally accessed within the 4-$f$ system, where a transmission mask is implemented. This mask, composed of two holes, cuts any light but the zero-order spot and one of the 1st-order diffraction spots. Importantly, the diameter of the hole cropping the zero-order has to be small enough to act as a low-pass filter to transform the transmitted light into a plane wave, acting as the reference beam just like in DHM. At the image plane of the 4-$f$ system, where the camera is positioned, one beam is coming from the 1st order, representing the image impinging on the camera with a tilt angle, and another beam is coming from the zero-order, acting as a reference plane wave. The retrieval algorithm is thus exactly the same as in DHM (Eq. \eqref{eqDHM}). DPM is called a common-path QPM technique, because no external reference arm exists. Here lies the interest of DPM and common-path techniques, discarding a separate reference arm that is a source of instabilities.

{\revision \subsection{Illumination-based differential phase contrast (DPC)}

In microscopy, the term DPC regroup different techniques that modify either the illumination or the detection angles to increase the contrast and retrieve quantitative information. In this paper, we focus on the most common implementation in QPM, that is DPC based on the sequential acquisition of a set of intensity images, using a standard bright field microscope, with various illumination conditions (illumination-based DPC). For this purpose, a DPC microscope involves a large numerical aperture in illumination (typically 0.25 to 0.4), created by an extended source of light located in the front focal plane of the condenser (or far enough from the sample in the case there is no K\"ohler illumination) \cite{OE23_11394,CB5_794}. This extended source is uniformly switched on in the case of bright field microscopy (Fig.~\ref{QPMsetups}(a)). In the case of DPC, the light source is cut into two parts about its center, named left and right (Fig.~\ref{QPMsetups}(e)). Each half of the illumination gives rise to an asymmetric illumination. The two corresponding images, $I_\mathrm{left}$ and $I_\mathrm{right}$, are acquired. Then, the light source is split about the orthogonal axis to generate two other illuminations and two new images that we name  $I_\mathrm{top}$ and $I_\mathrm{bottom}$. This set of 4 images are used to generate the DPC images defined by \cite{OE23_11394}
\begin{align}
I^\mathrm{DPC}_1&=\frac{I_\mathrm{top}-I_\mathrm{bottom}}{I_\mathrm{top}+I_\mathrm{bottom}},\label{eq:DPCexpressions}\\
I^\mathrm{DPC}_2&=\frac{I_\mathrm{right}-I_\mathrm{left}}{I_\mathrm{right}+I_\mathrm{left}}.
\end{align}
The basic idea of DPC is that these image subtractions contain information on the phase gradient, respectively along $x$ and $y$ directions. To demonstrate this relation, an important assumption needs to be made. First, one needs to consider the object as a plane of transmittance $t(x,y) = \exp(-\mu(x,y)+i\varphi(x,y))$, where $\phi$ is the phase image of interest, to be determined. Then, $\mu$ and $\phi$ have to be small enough so that
\begin{equation}
t\approx1-\mu+i\varphi.
\end{equation}
This approximation also neglects any cross term between $\mu$ and $\varphi$. In this condition, one can define some point-spread-functions in intensity and phase, namely $H_\mathrm{abs}$ and $H_\mathrm{pha}$ so that the intensity at the camera plane $I$ can be expressed using convolutions:
\begin{equation}
I = B + H_\mathrm{abs}\ast\mu + H_\mathrm{pha}\ast\varphi,\label{eq:I_DPC}
\end{equation}
where $B$ is the (uniform) intensity map at the image plane measured without object and $\ast$ denotes the convolution product. $H_\mathrm{abs}$ and $H_\mathrm{pha}$ are like point-spread-functions (PSFs) in intensity and phase. In this expression, considering an aberration-free system, only the last (phase) term depends on the illumination side (right or left, top or bottom). Here lies the interest of considering the image subtraction introduced above (Eqs. \eqref{eq:DPCexpressions}), which become:
\begin{align}
I^\mathrm{DPC}_1&=\frac{\left(H_\mathrm{pha}^\mathrm{top}-H_\mathrm{pha}^\mathrm{bottom}\right)\ast\varphi}{I_\mathrm{top}+I_\mathrm{bottom}},\label{eq:DPCexpressions2}\\
I^\mathrm{DPC}_2&=\frac{\left(H_\mathrm{pha}^\mathrm{right}-H_\mathrm{pha}^\mathrm{left}\right)\ast\varphi}{I_\mathrm{right}+I_\mathrm{left}}.
\end{align}
Note that the denominators of the two right-hand sides of the equation are supposed to be both equal to the bright field image. $\varphi$, the quantity of interest, is then retrieved by a deconvolution algorithm based on Fourier transforms and a Tikhonov's regularization:
\begin{equation}
\varphi = \mathcal{F}^{-1}\left[\frac{\sum_j\tilde H^*_j\tilde I^\mathrm{DPC}_j }{\sum_j\vert\tilde  H_j\vert^2+\epsilon}\right],\label{eq:DPCfinal}
\end{equation}
with $H_1=(H_\mathrm{pha}^\mathrm{top}-H_\mathrm{pha}^\mathrm{bottom})/B$, $H_2=(H_\mathrm{pha}^\mathrm{right}-H_\mathrm{pha}^\mathrm{left})/B$, where $\tilde A$ denotes the Fourier transform of $A$: $\tilde A=\mathcal{F}(A)$ and $\mathcal{F}^{-1}$ is the inverse Fourier transform. $\epsilon\ll1$ is the Tikhonov regularization parameter. Decreasing its value gives more accurate results but increases the noise level. $\epsilon=10^{-3}$ is considered as a good compromise \cite{OE23_11394,CB5_794}. $\epsilon$ is the important degree of freedom to consider in DPC, as it significantly affects the accuracy of the measurements, as explained hereinafter.

The aim of the article is to compare microscopy techniques on the exact same microscopes. However, DPC has never been used with high-NA objectives, because the NA of the objective has to match the NA of the illumination. For this reason, in the DPC simulations, as an exception, we considered an objective NA of 0.4, instead of 1.3.}

\subsection{Phase-shifting techniques \label{sect:PSI}}

With phase-shifting QPM techniques, two beams interfere at the image plane, but without tilt. They propagate collinearly so that no fringe pattern is produced. This approach, which looks more natural and simpler than the off-axis approach, does not offer the possibility to retrieve the phase from a single image acquisition. The missing information can be retrieved by acquiring 4 images with $0$, $\pi/2$, $\pi$ and $3\pi/2$ overall phase shifts applied to one of the two beams. Only with such 4 images can the phase profile be reconstructed. In the following, we described the 3 phase-shifting techniques presented in this article, namely PSI, FPM and SLIM.

\subsubsection{Phase-shifting interferometry (PSI) microscopy}

PSI is an old technique \cite{AO27_5082,AOP7_1,JM179_11,EPJAP44_29,  Book_Popescu}. In PSI, a reference plane wave $E_\mathrm{ref}$ is overlapped with the wave crossing the object plane (Fig.~\ref{QPMsetups}(f)), just like in DHM (Fig.~\ref{QPMsetups}(c)). However, no tilt is applied so that the total intensity reads 
\begin{align}
I(\mathbf{r})=&\vert E(\mathbf{r})+E_\mathrm{ref}\vert^2\nonumber\\
I(\mathbf{r})=&\vert E(\mathbf{r})\vert ^2+\vert E_\mathrm{ref}\vert^2\nonumber\\
&+2\vert E(\mathbf{r})\vert\vert E_\mathrm{ref}\vert\cos\varphi(\mathbf{r}).\label{eqPhaseShifting}
\end{align}
The issue with Eq. \eqref{eqPhaseShifting} is that the phase $\varphi$ can no longer be retrieved from a demodulation around a carrier spatial frequency. There is no bijective relation between $I$ and $\varphi$ in Eq. \eqref{eqPhaseShifting}, because $I$ also depends on $\vert E\vert$, which is usually non-uniform.

To lift this limitation, not one, but 4 images are acquired, $I_{j\in\llbracket0,3\rrbracket}$, corresponding to 4 different overall phase shifts $\phi_j$ between the object and reference fields, namely $0$, $\pi/2$, $\pi$ and $3\pi/2$:
\begin{align}
I_j(\mathbf{r})=&\vert E(\mathbf{r})+E_\mathrm{ref}\exp(i\phi_j)\vert^2\nonumber\\
I_j(\mathbf{r})=&\vert E(\mathbf{r})\vert^2+\vert E_\mathrm{ref}\vert^2\nonumber\\
&+2\vert E(\mathbf{r})\vert\vert E_\mathrm{ref}\vert\cos(\varphi(\mathbf{r})-\phi_j),\\
\mathrm{where}\;& \phi_j=j\pi/2, j\in\llbracket0,3\rrbracket\nonumber.
\label{eqPSI}
\end{align}
The four images $I_j$ enable the retrieval of the phase map $\varphi$ using the simple expression \cite{JM179_11}:
\begin{equation}
\varphi=\arg\big[(I_0-I_2)+i(I_1-I_3)\big].\label{eqvarphi}
\end{equation}

\subsubsection{Fourier phase microscopy (FPM)}

FPM is a phase-shifting technique that does not involve an external reference beam \cite{OL29_2503,AO46_1836}. This common-path technique only requires a simple plane wave illumination of the sample via a coherent light source, or a well-adjusted K\"ohler device, making this technique adaptable to any optical microscope without modifying it. Moreover, as a common-path technique, no temporal coherence is required and incoherent light sources can be used a priori. The two beams that interfere with artificial phase shifts are no longer $E$ and $E_\mathrm{ref}$, but $E_0$ and $E_\mathrm{s}$, {\it i.e.}, the incident and scattered fields (see Eq. \eqref{EE0Es}). By varying the phase-shift $\phi_j$ between $E_0$ and $E_s$ (see below), multiple images $I_j$ can be obtained:
\begin{align}
I_j(\mathbf{r})=&\vert E_0\exp(i\phi_j)+E_\mathrm{s}(\mathbf{r})\vert^2\nonumber\\
I_j(\mathbf{r})=&\vert E_0\vert^2+\vert E_\mathrm{s}(\mathbf{r})\vert ^2\nonumber\\
&+2\vert E_\mathrm{s}(\mathbf{r})\vert\vert E_0\vert\cos(\varphi_\mathrm{s}(\mathbf{r})-\phi_j).
\label{eqFPM}
\end{align}
   
Then, Eq. \eqref{eqvarphi} is used to retrieve $\varphi_\mathrm{s}$:
\begin{equation}
\varphi_\mathrm{s}=\arg\big[(I_0-I_2)+i(I_1-I_3)\big].\label{eqvarphis}
\end{equation}
However, $\varphi_\mathrm{s}$ is the phase map of the scattered field $E_\mathrm{s}$, not the phase map $\varphi$ of the total field $E$. With FPM, one needs another step to retrieve $\varphi$ compared with PSI, which is given by this expression:
\begin{equation}
\varphi(\mathbf{r})=\mathrm{arctan}\left(\frac{\beta(\mathbf{r})\sin\varphi_\mathrm{s}(\mathbf{r})}{1+\beta(\mathbf{r})\cos\varphi_\mathrm{s}(\mathbf{r})}\right),\label{phibeta}
\end{equation}
where the $\beta$ image is defined by $\beta=\vert E_\mathrm{s}/E_0\vert$ and can also be calculated from the 4 $I_j$ images:
\begin{equation}
\beta=\frac{1}{4E_0}\frac{I_0-I_2+I_3-I_1}{\sin(\varphi_s) + \cos(\varphi_s)}.\label{betaMap}
\end{equation}
Experimentally, while it was straightforward to specifically apply the 4 phase shifts on the reference beam in PSI, one has here to apply the phase shifts only to the scattered field $E_\mathrm{s}$, which is challenging because $E_\mathrm{s}$ and $E_0$ follow the same path. The assumption in FPM is to consider these two fields separated in the Fourier plane of the microscope. Indeed, the center of the Fourier space is where the light from $E_0$ is focused while the rest of the Fourier space contains only information related to the scattered field $E_\mathrm{s}$. Thus, it suffices to apply a phase mask over a tiny area at the center of the Fourier plane to specifically apply the $\phi_j$ shifts to $E_0$, which is performed using a spatial light modulator (SLM) conjugated with the back focal plane of the objective (Fig.~\ref{QPMsetups}(g)). One will see in this article that this assumption is not always valid because some scattered field can be contained within the central spot of the Fourier plane, in some cases. Moreover, the use of an SLM demands the use of a linearly polarised light beam.

\subsubsection{Spatial light interference microscopy (SLIM)}

SLIM \cite{OE19_1016} works exactly like FPM. There is just a slight difference with the experimental implementation. SLIM is meant to be adapted on a Zernike phase contrast (PC) microscope, involving a ring-like illumination and a special objective lens containing an annular phase mask. The annular mask of the objective is aimed to apply a phase shift of $\pi/2$ to the incident light field. The idea of SLIM is that PC microscopy can become quantitative if the phase shift of the annular mask could be varied and successively set to 4 values, to turn it into a phase-shifting technique. This is made possible by conjugating an SLM to the back focal plane of a PC microscope, applying annular phase shifts matching the geometry of the phase mask of the objective lens. Moreover, to avoid losing too much intensity with a beam splitter, SLIM benefits from a slightly shifted beam propagation before and after the SLM (Fig.~\ref{QPMsetups}(h)). Here again, we shall see hereinafter than considering that the annulus contains only information related to the incident light beam $E_0$ is too strong an approximation in most cases.

\subsection{Transport-of-intensity equation (TIE) microscopy\label{TIEalgoSection}}

{\revision TIE microscopy only requires a normal bright-field microscope. The wavefront information can be retrieved using a set 3 images captured at 3 different focuses of the sample $z$ (Fig.~\ref{QPMsetups}(f)). Here is the wavefront retrieval procedure.}

From the Helmholtz equation governing the propagation of a light wave $A(\mathbf{r})e^{i\varphi(\mathbf{r})}$, one can derive the so-called transport-of-intensity equation (TIE), which no longer involves the electric field amplitude $A$, but the light intensity $I = A^2$ \cite{OC199_65}:
\begin{equation}
\nabla_\perp\cdot\left[I(\mathbf{r})\nabla_\perp W(\mathbf{r})\right]=-\partial_zI(\mathbf{r}).
\label{eq:TIE}
\end{equation}
This equation is approximate, assuming that the light wave is a perturbation from a plane wave travelling along $z$. This equation tells how the transverse gradient of a light beam affects the variations of the intensity along the propagation axis. Assuming $I$ and $\partial_zI$ are known, this equation has a unique solution for the wavefront profile $W$, provided that $I>0$ everywhere in the image plane. To solve this equation, several methods exist. We use here the method proposed by Teague \cite{JOSA73_1434,OC199_65}, who showed that the vector field $I(\mathbf{r})\nabla_\perp W(\mathbf{r})$ derives from a scalar potential $V$:
\begin{equation}
I(\mathbf{r})\nabla_\perp W(\mathbf{r})=\nabla_\perp V(\mathbf{r}).
\label{eq:IWV}
\end{equation}
Injecting this expression in Eq. \eqref{eq:TIE} gives
\begin{equation}
\nabla_\perp^2 V(\mathbf{r})=-\partial_zI(\mathbf{r}).
\label{eq:nablaV}
\end{equation}
This equation in $V$ can be solved using a Green's function formalism \cite{JOSA73_1434}, or a Fourier transform formalism \cite{OC199_65}, which is the approach depicted below. The calculation of the gradient of a function can be performed using Fourier transforms:
\begin{align}
\nabla_\perp f(x,y)=&i\mathbf{u}_x\mathcal{F}^{-1}\left[q_x\mathcal{F}\left[f(x,y)\right]\right]\\
+&i\mathbf{u}_y\mathcal{F}^{-1}\left[q_y\mathcal{F}\left[f(x,y)\right]\right],
\end{align}
where $q_{x,y}$ are the variables conjugate to $x$ and $y$ in the Fourier space. Using this transformation, Eq. \eqref{eq:nablaV} can be solved
\begin{equation}
V(\mathbf{r})=\mathcal{F}^{-1}\left[q^{-2}\mathcal{F}\left[\partial_zI(\mathbf{r})\right]\right],
\label{eq:nablaV2}
\end{equation}
where $q^2=q_x^2+q_y^2$.
Finally, injecting this expression of $V$ in Eq. \eqref{eq:IWV} gives
\begin{equation}
W(\mathbf{r})=-\mathcal{F}^{-1}\left[q^{-2}\mathcal{F}\left[\nabla_\perp\cdot\left(I(\mathbf{r})^{-1}\nabla_\perp V(\mathbf{r})\right)\right]\right].\label{W_TIE}
\end{equation}

In practice, in TIE microscopy, a conventional wide-field, transmission microscope is used. Three intensity images are acquired at 3 different focuses $z=0$, $-\Delta z$, and $\Delta z$. Let us name them respectively $I_0$, $I_\mathrm{d}$, and $I_\mathrm{u}$. The out-of-focus intensity images enable the estimation of the image of the intensity $z$-gradient, at $z=0$:
\begin{equation}
\partial_zI=(I_\mathrm{u}-I_\mathrm{d})/2\Delta z.\label{deltatzI}
\end{equation}
This image is used to compute $V$ using Eq.~\eqref{eq:nablaV2}, and then $I_0$ is used in Eq.~\eqref{W_TIE} to compute the wavefront profile $W$. Note that with this approach, the intensity image is directly measured and given by $I_0$, without image processing, unlike with CGM/QLSI or DHM, for instance, that require a demodulation, implying a loss in image definition ({\it i.e.} number of pixels).

\subsection{Comparison of the QPM setups}

\begin{figure}[t]
	\centering
		\includegraphics[scale=0.87]{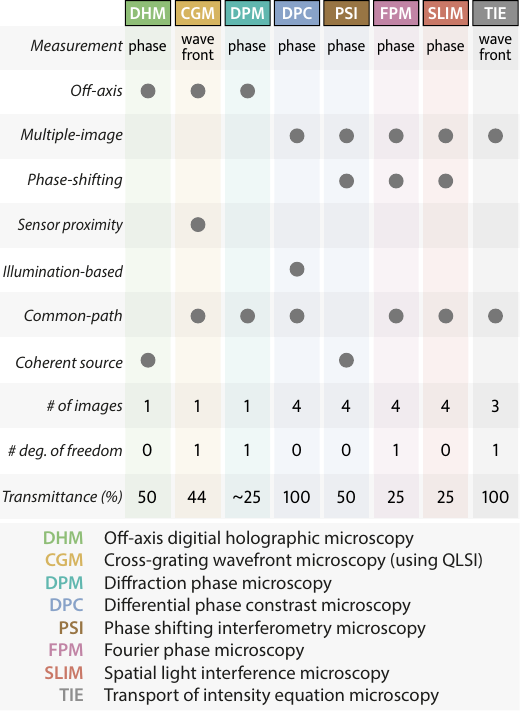}
	\caption{{\bfseries Technical features of the QPM techniques presented in this study}, namely what they measure (wavefront or phase), their category (off-axis, phase-shifting, phase-diversity, or common-path), if they require a coherent source of light ({\it i.e.} cw laser illumination), the typical number of images required to form a phase/wavefront image, the number of experimental degrees of freedom, and the transmittance of the setup. The list of the abbreviations and their meanings are give below the table.}
	\label{QPMcategories}
\end{figure}

Figure~\ref{QPMcategories} summarizes the 8 QPM techniques presented in this study, their relationship to QPM categories (off-axis, multiple-image, phase-shifting, etc). Some features are also listed (number of images, degrees of freedom, etc).

Some techniques require the use of a coherent source of light ({\it i.e.} a cw laser), in particular with techniques that use a reference arm that do not necessarily have the same length as the main arm (not common-path techniques), such as DHM or PSI. Although interferences are involved in CGM/QLSI, DPM, FPM and SLIM, these techniques do not need the use of coherent light sources because the beam path is the same between the different beams interfering at the image plane.

Figure~\ref{QPMcategories} also displays the number of images that need to be acquired to construct the phase/wavefront images. The higher this number, the slower the technique. In PSI, FPM and SLIM, although 3 images would be sufficient in theory, 4 images are usually acquired to substantially improve the signal-to-noise ratio.

Some QPM techniques possess a degree of freedom (DoF) that can be adjusted experimentally. In CGM/QLSI, it is the grating-camera distance. In DPM, the DoF is the diameter of the 0-order hole in the mask. In DPC, the numerical aperture of the illumination has to match the NA of the objective lens, so the DoF is not really a DoF. DPC rather possesses a \emph{numerical} DoF, which is the Tikhonov regularization parameter $\alpha$, and which can affect both the trueness and precision of the measurements. In FPM, the DoF is the diameter of the disc area on the SLM that phase-shifts the central part of the Fourier space. In TIE, the DoF is the defocus $z$ between successive acquired images. As will be explained later, these DoFs are associated with a trade-off between trueness and precision.

The last line of this table corresponds to the optical transmission of the QPM techniques, defined as the $t/t_0$ where $t$ is the percentage of photons reaching the camera compared with the number of photons impinging on the field of view of the microscope, and $t_0$ is $t$ for a standard, bright-field transmission microscope (see Fig.~\ref{QPMsetups}). For CGM/QLSI, the transmission is the one of the grating, in which the opaque lines cover 5/9 of the area, leading to a transmission of  44\%. DHM loses 50\% due to the beam splitter. DPM loses 50\% because of the opaque lines of the diffraction grating, and several 10\% more due to the pinhole. PSI loses 50\% because of the beam splitter. FPM transmission is 25\% because the signal passes twice through a beam splitter. The transmission of SLIM is 50\% because of the use of a linear polarizer to ensure a linear polarisation on the SLM~\cite{AOP13_353}, and 50\% more because the phase mask in a phase-contrast objective lens usually also absorbs 50\% of the intensity to improve the contrast. The transmission of DPC and TIE is 100\% because they use a conventional bright-field microscope, with no additional optical element.

\section{Numerical simulations using IF-DDA}

Comparing 8 different experimental techniques with each other would have been cumbersome experimentally. Some articles experimentally compare several QPM techniques, but never more than 3 techniques \cite{OE26_17498,BOE10_2768,OL39_5511}, and not always on the same microscope with the same cameras, and sometimes not even using the same magnifications and numerical apertures~\cite{BOE10_2768}, making any objective comparison complicated.

Here, we rather chose to \emph{numerically} compare QPM techniques, so that we are not limited by the number of techniques, and we can compare them on strictly equivalent microscopes, on strictly identical objects.

The numerical tool we use to model the microscopy techniques is based on the discrete dipole approximation (DDA) method, a common approach for electromagnetism simulations~\cite{Draine_AJ_88}. DDA algorithms require the meshing of only the object, not of the surrounding medium, and can easily take into account the presence of a flat interface (on which the sample is lying)~\cite{Chaumet_M_22}. We recently released a home-made DDA toolbox called IFDDA (for Institut-Fresnel or Idiot-Friendly DDA)~\cite{JOSAA38_1841,IFDDAgitlab}, which not only models the interaction between a light beam and an object, but also computes the image of this object by a microscope~\cite{JOSAA38_1841,O7_243}, {\it i.e.}, the electromagnetic field at the image plane. {\revision It should be noted that all the calculations performed by the IFDDA code are made without any approximation within the vectorial framework.}

\begin{figure*}
	\centering
        \rotatebox[origin=c]{90}{\includegraphics[scale=0.89,]
        {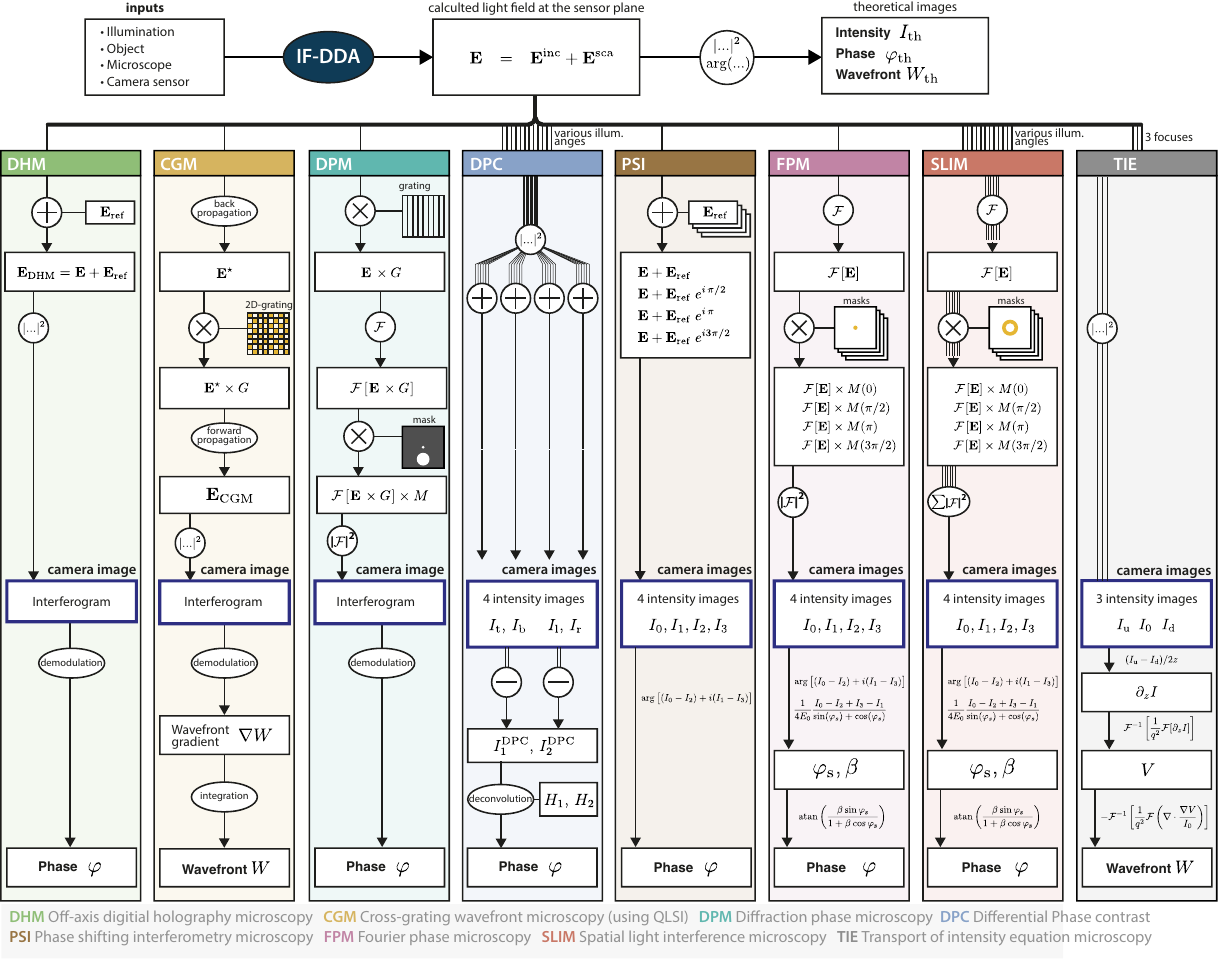}}
	\caption{{\bfseries Computing chart of \emph{in silico} quantitative phase microscopies using IF-DDA.} This chart shows the numerical procedure, starting from the computation of the light field at the sample plane of the microscope, and its processing using 8 different QPM microscope models, until the computation of the experimental-like intensity, phase, and wavefront images. These modelled images are aimed to be compared with the theoretical intensity, phase, and wavefront images directly calculated from the theoretical light field, as a means to evidence inherent artefacts of specific QPM, or quantify the noise amplitude on the modelled images, coming from the shot noise on the camera sensor.}
	\label{codeChart}
\end{figure*}

In all numerical simulations, the incident illumination took the form of a plane wave or a superposition of plane waves, characterized by intensity, polarization (linear or circular), direction (not necessarily at normal incidence), and wavelength. The modeled microscope in all simulations involved a $100\times$ magnification, 1.3 NA objective lens, and a camera with a pixel size of 6.5 \textmu m and a full-well-capacity of 25,000. The only exceptions were the use of a $60\times$ magnification exclusively for the neuron, and a 0.4 NA objective only for DPC simulations.

The program returns the vectorial electric field of the light at the image plane $\mathbf{E}$, being composed of the incident light $\mathbf{E}_0$ and the light scattered by the object $\mathbf{E}_\mathrm{s}$ (Eq.~\eqref{EE0Es}). From $\mathbf{E}$, we compute the theoretical intensity $I_\mathrm{th}=\vert\mathbf{E}\vert^2$, phase $\varphi_\mathrm{th}$ and wavefront $W_\mathrm{th}$ images at the image plane of a conventional (bright-field) microscope (see upper part of Fig.~\ref{codeChart}).

While computing $I_\mathrm{th}$ from $\mathbf{E}$ is straightforward, computing $\varphi_\mathrm{th}$ or $W_\mathrm{th}$ is less obvious. First, as explained above, computing the phase and wavefront profiles of an electromagnetic field is not straightforward when taking into account polarisation: There are actually two phase maps, one for each transverse component of the light beam:
\begin{align}
\varphi_x=& \arg(E_x)\\
\varphi_y=& \arg(E_y).
\end{align}
In all the simulations we conducted, the scalar approximation is supposed to remain valid, because none of the objects are birefringent. Nevertheless, in the simulations, we consider that the true phase map equals the average of these two profiles, weighted by the intensity maps:
\begin{align}
\varphi_\mathrm{th}=\frac{I_x\varphi_x+I_y\varphi_y}{I_x+I_y}\label{phiWeighted}.
\end{align}
This expression was derived in the case of CGM experiments \cite{OE23_16383}, and we used it in this study for any QPM to determine the true phase profiles $\varphi_\mathrm{th}$, and to compute $W_\mathrm{th}=\varphi_\mathrm{th}\lambda/2\pi$.

In practice, $\varphi_\mathrm{th}$ can sometimes exceeds variations of $2\pi$. In that case, an unwrapping algorithm is necessary. For the simulations presented in this study, we did not need to use unwrapping algorithms, because all the objects were not optically thick enough.\\

In QPM, this initial electromagnetic field $\mathbf{E}$, originally impinging on the camera in bright field microscopy, is transformed by optical elements, such as gratings, masks, and polarizers. To conduct the simulations presented in this article, we upgraded the IF-DDA toolbox so that it no longer computes only $\mathbf{E}$, but also the transformations generated by any QPM. More precisely, we modeled all the optical elements for each QPM technique (gratings, polarizers, and masks), and computed the resulting electric field at the camera plane. The numerical procedures for all the QPMs are sketched in Fig.~\ref{codeChart}. The codes of IF-DDA and of the QPM add-on are provided on public repositories \cite{IFDDAgitlab}. This way, one can compute the raw image recorded by the camera for each QPM, and process it to get the phase image $\varphi$ or wavefront image $W$ using experimental processing algorithms, as if it were an actual experimental image. We can then compare the \emph{true} maps, $I_\mathrm{th}$, $\varphi_\mathrm{th}$ and $W_\mathrm{th}$, at the image plane and calculated using IF-DDA, with the so-called \emph{modelled} maps $I$, $\varphi$, and $W$ obtained by post-processing the raw image captured by the camera, for each specific QPM. This way, any inaccuracy inherent to the working principle of QPM techniques can be evidenced, quantified and compared.

For the techniques requiring a laser illumination or a linearly polarized illumination (see Fig.~\ref{QPMcategories}), we used a linearly polarized illumination. For the techniques using incoherent illumination, we used a circularly polarized illumination, to equally distribute the energy over the different polarisation axes.

Moreover, in the second part of our study, we numerically affected the raw camera image with shot noise as a means to quantify and compare the signal-to-noise ratios and precision of all the QPM techniques, assuming they are shot-noise limited. We neglect here other sources of noise, such as coherent noise (also called speckle noise), thermal and dark current noise.\\

Here are more details on this \emph{in silico} experimental approach, the algorithms of which are sketched in Fig.~\ref{codeChart}, technique by technique.

\subsection{CGM} 

To model CGM, we use the \emph{in silico} procedure that we recently used and detailed in Ref.~\cite{OC521_128577}. Briefly, we consider a circularly-polarised light beam impinging on the object at the sample plane at normal incidence, and compute the resulting electric field $\mathbf{E}$ at the image plane using IF-DDA.

Then, we back-propagate $\mathbf{E}$ using a Fourier-transform algorithm the grating-camera distance $d$. We multiply the back-propagated field by the complex transmittance of a QLSI grating, rotated by an angle of 53$^\circ$ around the optical axis $(Oz)$ (to minimize Moir\'e effects), and then forward-propagate over the same distance $d$ to get the interferogram recorded by the camera. Finally, we use an experimental processing algorithm to compute the intensity $I$ and wavefront image $W$ from the interferogram. Importantly, since CGM is not measuring a phase, but a wavefront gradient that is subsequently integrated, no unwrapping algorithm needs to be used in CGM. A CGM algorithm does not yield discontinuities on the phase/wavefront maps. The related Matlab code of this specific algorithm is provided on GitHub \cite{GitHub-CGMprocess,GitHub_CGMinSilico}.

\subsection{DHM} To model DHM, we consider a linearly-$x$-polarised light beam impinging on the object at the sample plane at normal incidence, and compute the resulting electric field $\mathbf{E}$ at the image plane using IF-DDA. Then, at the image plane, we add the reference beam, as a linearly polarised plane wave tilted by an angle $\theta$ around the $(Ox)$ axis: $\mathbf{E}_\mathrm{ref}=E_\mathrm{ref}\exp(i\mathbf{k}\cdot\mathbf{r})\mathbf{u}_x$ where $\mathbf{k}=(0,k\sin\theta,k\cos\theta)$. The intensity recorded by the camera is then simply calculated as
\begin{equation}
I=\vert\mathbf{E}+\mathbf{E}_\mathrm{ref}\vert^2.
\end{equation}
Then, the intensity map $I$ is processed using a DHM algorithm, as if it were an experimental image, to compute the \emph{modelled} phase map $\varphi$.

\subsection{DPM} To model DPM, we consider a circularly-polarised light beam impinging on the object at the sample plane at normal incidence, and compute the resulting electric field $\mathbf{E}$ at the image plane.

Then, we multiply $\mathbf{E}$ by the transmittance of the 1D grating rotated by 45$^\circ$ around the optical axis $(Oz)$. The pitch of the grating is such that it corresponds to 3 dexels (pixels of the detector). The resulting field is then Fourier-transformed and multiplied by the DPM mask passing the zero order and one 1st order. We shall call $D_0$ the diameter of the circular crop of the zero order, and $D_1$ the diameter of the circular crop of 1st order. We considered that the 4-$f$ system was composed of two lenses of focal lengths 10 cm. Finally, the cropped field is back-Fourier-transformed to get the interferogram recorded by the camera. This interferogram is processed using a DPM algorithm to retrieve the phase image $\varphi$, which is exactly the same as the DHM algorithm.

\subsection{DPC} To model DPC and its 4 tilted illuminations, we reproduced numerically what is usually done experimentally: using an array of point-like sources of light (like a LED array). We chose a circular distribution of sources, with a square unit cell, and with 16 sources along a diameter, leading to 192 sources in total (see the LED array geometry in Fig. S1 in Suppl. Info.). The size of the pattern was adjusted so that it produces a numerical aperture of 0.4 in illumination. We ran a series of 192 numerical simulations, corresponding to all these incident plane waves, all circularly polarized. The single, true phase map $\varphi_\mathrm{th}$ is assumed to be the phase map for normal incident illumination at zero NA. For each numerical simulation $j$, the intensity map $I_j$ at the camera plane is calculated. Then, these intensity maps are grouped, depending on which side they belong to (top, right, left or bottom, see Fig. S1 in Suppl. Info.), and summed to get the 4 DPC intensity images $I_\mathrm{top}$,  $I_\mathrm{right}$,  $I_\mathrm{left}$ and  $I_\mathrm{bottom}$, from which the 2 contrast images $I^\mathrm{DPC}_1$ and $I^\mathrm{DPC}_2$ are calculated, using Eqs.~\eqref{eq:DPCexpressions}. The intensity and phase PSFs are calculated using the expressions given in Suppl. Info. Then, the phase map is calculated using Eq. \eqref{eq:DPCfinal}.

\subsection{PSI} To model PSI, we consider a linearly-$x$-polarised light beam impinging on the object at the sample plane at normal incidence, and compute the resulting electric field $\mathbf{E}$ at the image plane using IF-DDA.

Then, at the image plane, we add the reference beam $\mathbf{E}_\mathrm{ref}$, as a linearly $x$-polarised plane wave at normal incidence endowed with a given, uniform phase shift $\phi$: $\mathbf{E}_\mathrm{ref}=E_\mathrm{ref}e^{i\phi}\mathbf{u}_x$. The intensity recorded by the camera is then simply calculated as $I=\vert\mathbf{E}+\mathbf{E}_\mathrm{ref}\vert^2$. 4 images are generated, corresponding to 4 different phase shifts $\phi$ of $0$, $\pi/2$, $\pi$ and $3\pi/2$. These 4 images are then used to compute the phase image $\varphi$ using Eq.~\eqref{eqvarphi}.

\subsection{FPM} To model FPM, we consider a linearly-$x$-polarised light beam impinging on the object at the sample plane at normal incidence, and compute the resulting electric field $\mathbf{E}$ at the image plane using IF-DDA. Then, the field is Fourier-transformed and a phase $\phi$ is applied at the center of the Fourier space, over a circular area. The diameter $D_0$ of this circular area captures light over an area corresponding to a numerical $\mathsf{NA}_0=D_0M/f$ in the object space, where $f$ is the focal length of the tube lens. We then inverse-Fourier-transform the field to get the intensity image recorded by the camera. 4 images are generated, corresponding to 4 different phase shifts $\phi$ of $0$, $\pi/2$, $\pi$ and $3\pi/2$. These 4 images are then used to compute the phase $\varphi_\mathrm{s}$ of the scattered field using Eq.~\eqref{eqvarphi}, which is then used to compute the phase image $\varphi$ of the total field using Eq. \eqref{phibeta}.

To model a non-zero illumination numerical aperture ($\mathsf{NA}_\mathrm{ill}=0.02$), we consider an ensemble of 13 plane waves corresponding to 13 point sources regularly spaced in the Fourier plane of the illumination ({\it i.e.}, at the aperture diaphragm location of the K\"ohler illumination);

\subsection{SLIM} To model the annular illumination in SLIM, we run a series of 12 numerical simulations, corresponding to various incident plane waves, all circularly polarized, regularly distributed along a ring around the optical axis $(Oz)$ at an NA of 0.25. For each of the 12 simulations, the true phase map is calculated as the phase map referenced by the phase map of the incident field $E_0$ at the sample plane. The single, true phase map $\varphi_\mathrm{th}$ is assumed to be the average of the 12 phase maps. For each numerical simulation, the $\mathbf{E}$ field is computed, Fourier-transformed, and a phase mask is applied. It consists of an annulus mask, of phase $\phi$ and transmittance $\tau=50$\% (the common transmittance of the phase ring of a phase contrast objective). The radius of the annulus matches the NA of the ring illumination (0.25), and its width is set to a NA of 0.02. The field is then inverse-Fourier-transformed to get the intensity at the camera plane. The intensities of all the 12 illuminations are summed (incoherent summation) to get the image recorded by the camera. 4 of such images are generated corresponding to the 4 values of $\phi$: $0$, $\pi/2$, $\pi$ and $3\pi/2$. These 4 images are then used to compute the phase $\varphi_\mathrm{s}$ of the scattered field using Eq. \eqref{eqvarphi} and the $\beta$ map using Eq. \eqref{betaMap}, which are then used to compute the phase image $\varphi$ of the total field using Eq. \eqref{phibeta}. Note that for SLIM, Eq. \eqref{betaMap} must be corrected to take into account the transmittance $\tau$ of the phase ring of the objective: $\beta=\sqrt{\tau}\vert E_\mathrm{s}/E_0\vert$.

\subsection{TIE}
To model TIE, we simulated three intensity images, $I_d$, $I_0$, $I_u$ of the sample at 3 different focus values of $-\Delta z$, 0 and $\Delta z$. Then, the three images were used to process the modelled intensity and wavefront images using the algorithm depicted in Sect. \ref{TIEalgoSection}.

\subsection{Noise simulation}
We also studied the presence of noise in OPD images. Generally, in QPM, the origin of noise is the shot noise, rather than the reading or thermal noise, because the camera sensor is usually well exposed. Shot noise comes from the discrete nature of light energy (photons) that produces a standard deviation in each dexel equal to $\sqrt{N}$, where $N$ is the number of photons collected by the dexel. Consequently, the larger the full-well-capacity of the camera, the better the signal-to-noise ratio \cite{OL41_1656,S_2304564,OC521_128577}. For the simulations intended to study the effect of shot noise, we generated noise of the raw camera images using the \texttt{poissrnd} function of Matlab, considering a full-well-capacity of the camera of 25,000.

The only source of noise simulated in this article is the shot-noise, inherently present for all techniques. We are aware that other sources of noise can be dominant, especially the coherent noise for techniques requiring laser illumination We did not simulate this type of noise, but discuss it further on.

To quantitatively compare the noise level of all the QPM techniques for a given object, the object has to be imaged with the same microscope settings (magnification, numerical aperture), and in the same conditions of illumination (at least the wavelength), which is the easy part. The more subtle aspect involves determining the quantity of light collected by the systems. This parameter is crucial as it directly influences the amplitude of noise in the images. It is more subtle because one can imagine different reference conditions, all equally justifiable, namely:
\begin{enumerate}[label=(\roman*)]
\item Camera full-well-capacity filled in any acquired image, no matter how many images the QPM needs to acquire, 1, 3 or 4 (which corresponds to what is done experimentally, most of the time).
\item Equal number of photons detected by the camera, and the illumination intensity is adjusted accordingly for each QPM (to discard the differences in optical transmissions of the QPMs).
\item Equal number of photons impinging on the sample (makes sense when the limitation is cell phototoxicity for instance)
\end{enumerate}
These three conventions are not equivalent. Convention (i) is not equivalent to (ii) because phase shifting techniques and TIE require the acquisition of several images (resp. 4 and 3), not just one. Then, (ii) is not equivalent to (iii) because QPMs have different transmissions (see Fig.~\ref{QPMcategories}). By convention, in Fig.~\ref{resultsNoise}, we chose Convention (ii), because Convention (i) would favor phase-shifting techniques that acquire 4 images instead of one, {\it i.e.}, 4 times more photons leading to a factor of 2 of noise reduction; and not Convention (iii) because phototoxicity is usually not an issue in QPM.\\

\section{Results\label{results}}
\begin{figure}[b]
\centering
\includegraphics[scale=0.9]{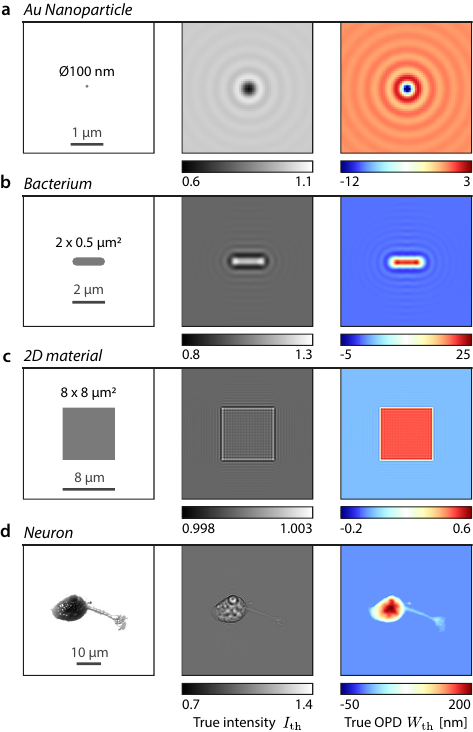}
\caption{{\bfseries The 4 objects investigated in this study}. (a) The first column defines their geometry, the second displays the theoretical intensity images and the third column displays the theoretical OPD (or wavefront) images. }
\label{objects}
\end{figure}

\begin{figure*}
\centering
\includegraphics[scale=0.9]{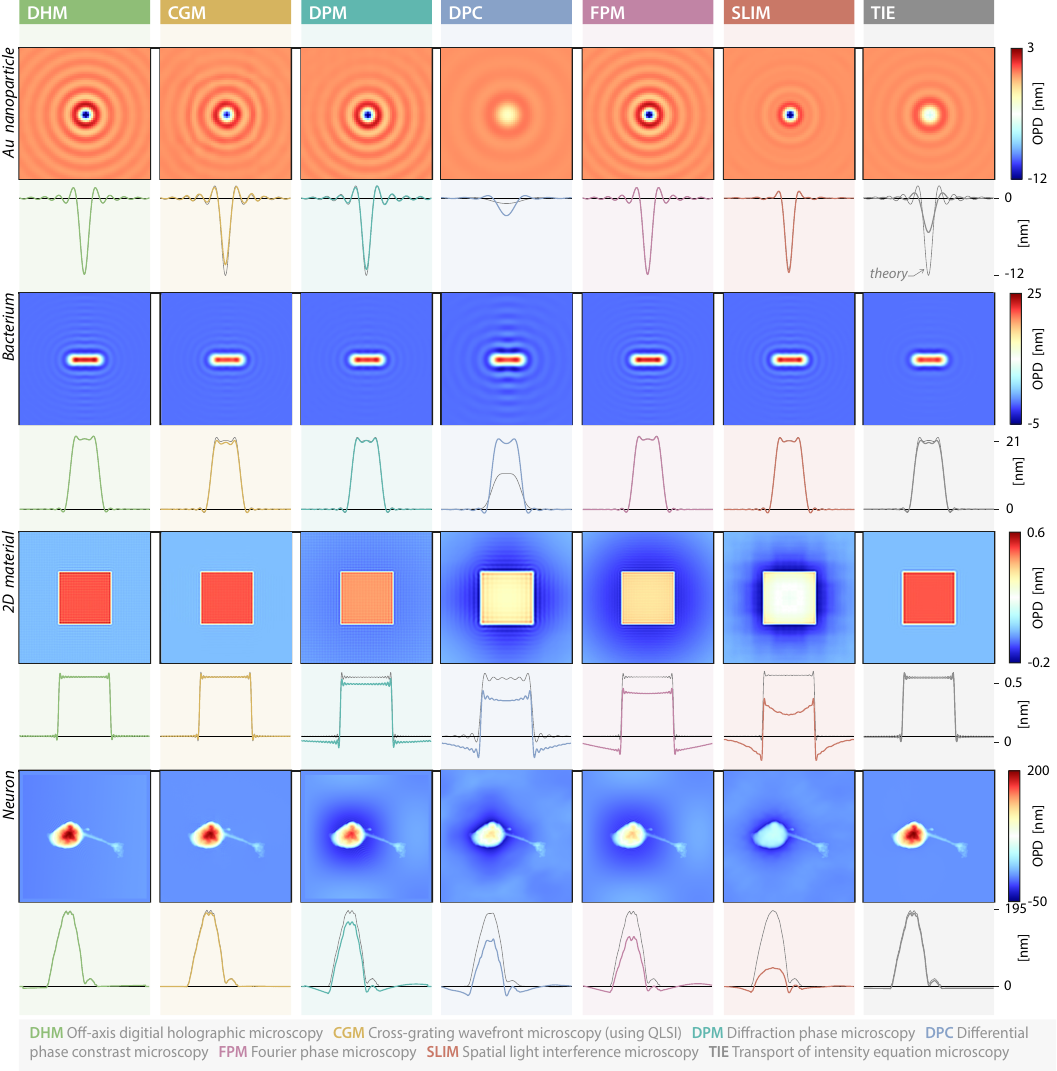}
\caption{{\bfseries Investigation of QPM trueness}. 28 numerical simulations of OPD images of 4 different objects, acquired using 7 different wavefront and phase microscopy techniques, on the same microscope. Each column corresponds to a microscopy technique. Each row corresponds to an object. For each image,  a horizontal crosscut passing through the center of the image is displayed below. The grey lines in the background indicate the theoretical profiles. The experimental setting parameters are: $d=0.6$ mm for CGM, $D_0=20$ \textmu m for DPM, $\epsilon=10^{-3}$ for DPC, $\mathsf{NA}_0=0.02$ for FPM, $\Delta z=500$ nm for TIE. PSI results are presented in Suppl. Info.}
\label{resultsTrueness}
\end{figure*}

\begin{figure*}
	\centering
		\includegraphics[scale=0.9]{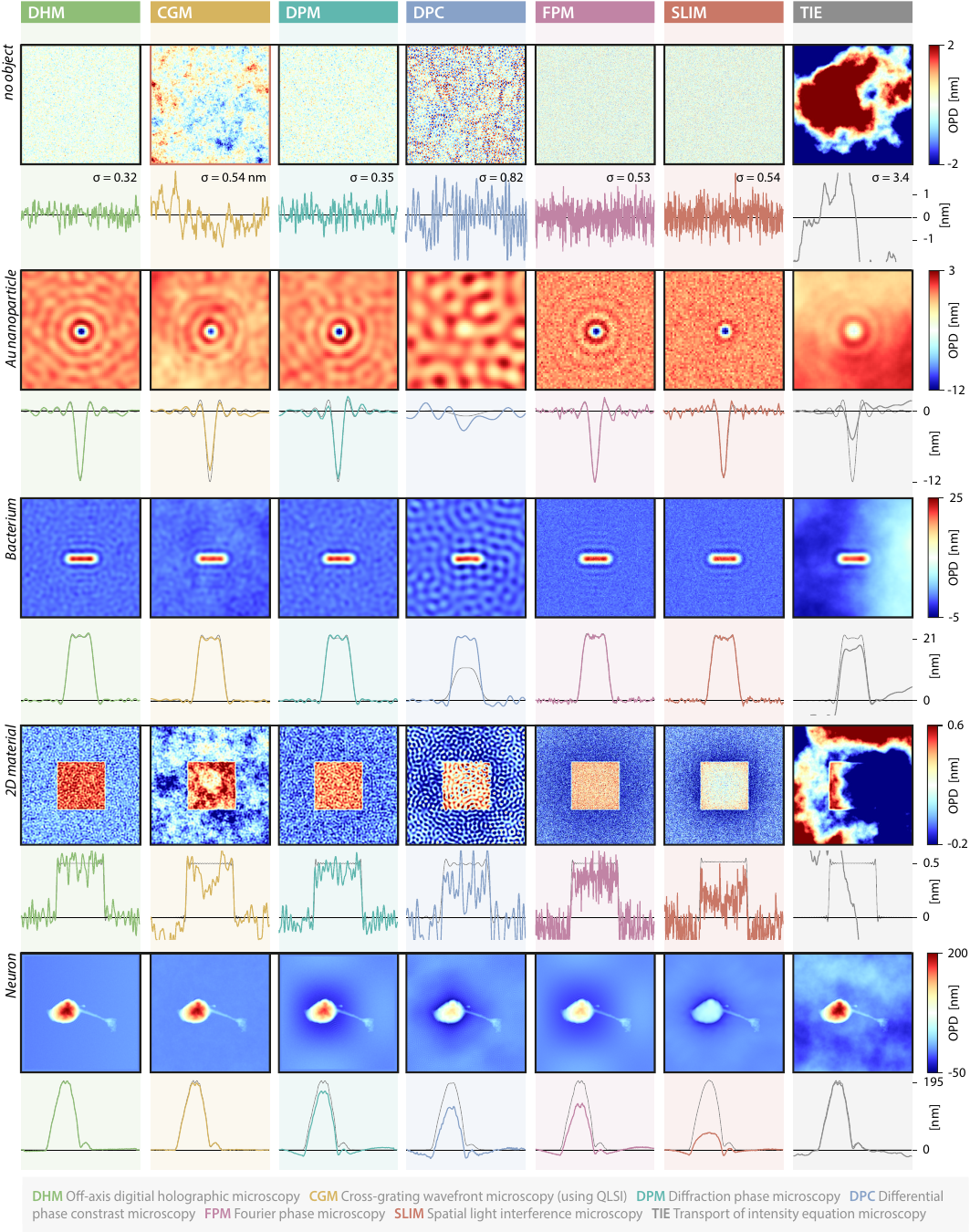}
	\caption{{\bfseries Investigation of QPM noise amplitudes}. 28 numerical simulations of OPD images of 4 different objects, acquired using 7 different wavefront and phase microscopy techniques, on the same microscope. Shot noise has been numerically added to each camera image. Each column corresponds to a microscopy technique. Each row corresponds to an object. For each image,  horizontal crosscuts passing through the center of the image are displayed below. The grey lines in the background indicate the theoretical profiles. In the first Row, $\sigma$ is the standard deviation on the images, in nm.  PSI results are presented in Suppl. Info.}
	\label{resultsNoise}
\end{figure*}

\begin{figure*}[!h]
	\centering
		\includegraphics[scale=0.9]{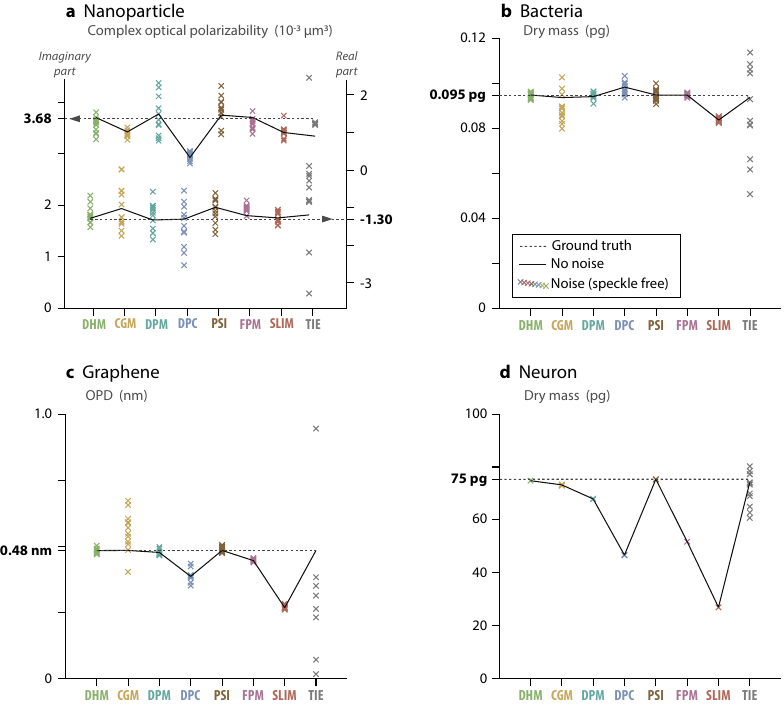}
	\caption{{\bfseries Biophysical quantities extracted from the simulations of Figs. \ref{resultsTrueness} and \ref{resultsNoise}.} (a) Plot of the complex optical polarisabilities of the 100-nm gold NP, measured with each technique. (b) Plot of the dry mass values of the bacteria. (c) Plot of the OPD at the 2D material. (d) Plot of the dry mass values of the neuron. Each graph represents ground truth as dashed lines, while solid lines connect values derived from noise-free images. Data points obtained from sets of 12 noisy (ideal, speckle-free) images are denoted by crosses.}
	\label{DMalphaMeas}
\end{figure*}

This part describes the results of the simulations related to the 8 QPM techniques (CGM, DHM, DPM, DPC, PSI, FPM, SLIM and TIE), imaging 4 different objects, namely a gold nanoparticle, a bacterium, a 2D material, and a neuron (Fig.~\ref{objects}). These objects have been chosen because they are commonly studied systems, and because they cover very different object dimensions. The gold nanoparticle (Fig.~\ref{objects}(a)) is spherical, 100 nm in diameter, and immersed in a uniform medium of refractive index 1.5, to match the experimental conditions of Ref.~\cite{O7_243}. The bacterium (Fig.~\ref{objects}(b)) is 2 \textmu m long, 0.5 \textmu m in diameter, lying on glass and immersed in water. It is endowed with a uniform refractive index of 1.38. The 2D material (Fig.~\ref{objects}(c)) is $8\times8$ \textmu m$^2$ in size, lying on glass, immersed in water, and designed to create an OPD of 0.5 nm, similar to the one of graphene \cite{ACSP4_3130}. The 3D model of the neuron (Fig.~\ref{objects}(d)) was created from an experimental image taken from Ref.~\cite{ACSP4_3130}. It was endowed with a uniform refractive index of 1.38 and the thickness of the cell was set to yield an OPD that was consistent with the experimental measurement. Considering a uniform refractive index within the cell is a strong approximation, but it is sufficient within the scope of this study to evidence the limitations of QPMs for such large objects. For the nanoparticle, bacterium and 2D material, the parameters of the microscope were $\lambda = 532$ nm, $100\times$ magnification, 1.3 NA. For the neuron, they were $\lambda = 600$~nm, $60\times$ magnification, 1.3 NA.\\

The aim here is to investigate the accuracy of all the QPMs. The accuracy is normally defined as the association of trueness and precision. Trueness refers to any systematic deviation of the modelled images from the true images, {\it i.e.}, bias or artefacts, while precision refers to inaccuracy measurements stemming from the noise on the images.

Figure~\ref{resultsTrueness} gathers numerical simulations of OPD images of the 4 objects imaged with 7 microscopy techniques, compared with the true OPD images (gray lines labelled 'theory'). Results on PSI are not included in the figures, for space reasons, but they are presented in Suppl. Info, Figure S2. In all these simulations, no shot noise was added to the raw camera images, in order to focus on the presence of possible artefacts.

Figure~\ref{resultsNoise} plots similar data, where shot noise has been added to the raw camera image. This figure is rather aimed to study measurement precision, in terms of signal-to-noise ratio. The 5 investigated systems are (i) a blank area, to better quantify the noise level, (ii) a 100-nm gold nanoparticle, (iii) a bacterium, (iv) a 2D material, and (v) a neuron. Note that for the 2D material, which features an OPD of only 0.5 nm, 25 images have been averaged for each technique to yield a more reasonable signal-to-noise ratio.

{\revision The results presented in Figs.~\ref{resultsTrueness} and~\ref{resultsNoise} highlight various levels of noise and deviations from ground truth among the different techniques. In order to quantify the degrees of accuracy of all the QPMs revealed by these simulations, we measured biophysical quantities from all these images. Specifically, we determined the dry mass values of the bacterium and the neuron, the complex optical polarisability of the NP and the average OPD of the 2D material. The dry mass values were evaluated by a pixel summation after tight segmentation of the cells, and the optical polarisability of the NP was determined according to the expression given in \cite{O7_243}. These values are plotted in Fig.~\ref{DMalphaMeas}. 

Measurements on the nanoparticle feature the largest lack of precision, being the smallest object (Fig.~\ref{DMalphaMeas}(a)). Even, on the noise-free images, the measurements vary from one technique to another, not really because of artefacts, rather because of the presence of diffraction rings making the segmentation more difficult. The nanoparticle does not feature significant artefacts, for any technique, as observed in Fig.~\ref{resultsTrueness}. Similarly, measurements on bacteria are not affected by artifacts, regardless of the QPM used (Fig.~\ref{DMalphaMeas}(b)). SLIM just features a weak underestimation. Measurements are mostly inaccurate with large objects, such as graphene (Fig.~\ref{DMalphaMeas}(c)) and eukaryotic cells (Fig.~\ref{DMalphaMeas}(d)); and with DPC, FPM and SLIM.

The presence of noise observed in Fig.~\ref{resultsNoise} makes the biophysical measurements more dispersed around the true value. To quantify this dispersion, we conducted a series of 12 identical simulations, in each QPM/object case, including random shot noise. The 12 measurements are also plotted in Fig.~\ref{DMalphaMeas}, so that this figure gives a general overview of the accuracy of all the techniques, in terms of trueness and precision, as a function of the imaged object. A larger noise affect CGM and TIE images in most cases, in particular for bacteria and graphene.

The following sections aim to explain the origins of all the QPM inaccuracies observed in Fig.~\ref{DMalphaMeas}. We  shall discuss the degrees of trueness (artefacts) and precision (signal-to-noise ratio) of each of the microscopy techniques addressed in this study, one by one.
}

\subsection{CGM accuracy: effect of the grating-camera distance\label{CGMaccuracy}}
\begin{figure*}
	\centering
		\includegraphics[scale=0.9]{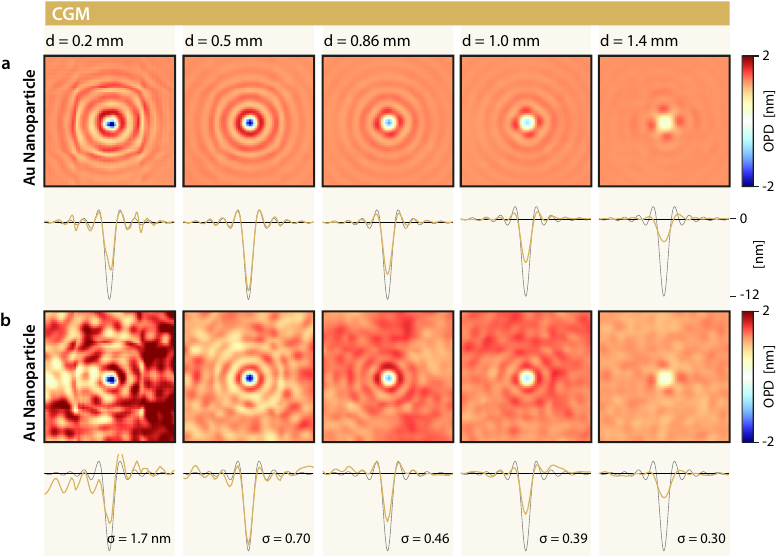}
	\caption{{\bfseries Influence of the grating-camera distance in cross-grating microscopy (CGM) on trueness and precision}. IF-DDA simulations of OPD images acquired with CGM, without (a) and with (b) noise. A cross-cut of the noisy images is displayed in the third raw. Each row corresponds to a different grating-camera distance, namely 0.2, 0.5, 0.86 and 1.2 mm. 0.86 mm corresponds to the common distance in CGM, measured on a commercial system.}
	\label{CGMgratingDistance}
\end{figure*}

All in all, the cross-cuts of the OPD images of CGM shown in Fig.~\ref{resultsTrueness} are in very good agreement with the theoretical profiles. Nevertheless, one can distinguish a slight discrepancy for the nanoparticle. The slightly reduced peak amplitude observed on the CGM image comes from the grating-camera distance (set at 0.6 mm in all the simulations presented in this Fig.~\ref{resultsTrueness}). The setting of the grating-camera distance is not critical, in the sense that there is no particular value that makes CGM work. The grating-camera distance $d$ can be continuously tuned, and the interferogram remains well-defined, without being blurred by a Talbot effect, thanks to the $0-\pi$ checkerboard pattern \cite{AO39_5715,JPDAP54_294002}. Figure~\ref{CGMgratingDistance}(a) shows CGM image simulations of the 100-nm gold nanoparticle for various grating-camera distances. The results show that the grating-camera distance must remain below a critical value to yield accurate OPD profiles. Above this limit, on the order of 1 mm for common QLSI systems in microscopy, the OPD amplitude is reduced until a point where 4-fold symmetry artefacts appear on the image \cite{OC521_128577}.\\

In CGM, the noise of the OPD images is a Flicker noise, also called a Brown noise. It is not a white noise like in most of the other QPM techniques. A Flicker noise is characterised by a larger amount of low spatial frequencies \cite{OC521_128577}, which makes it different from any other QPM technique. More precisely, the power spectral density of the noise in CGM scales as $1/f^2$, where $f$ represents the spatial frequencies of the image. This noise appears as the main drawback of CGM. Fortunately, it is weak, and can be removed by summing images (unlike a speckle noise). However, a Flicker noise is more difficult to remove using postprocessing than a white noise.

As observed in Fig.~\ref{CGMgratingDistance}(b), placing the grating further leads to a reduction of the noise amplitude on the OPD images. Moving the grating from 0.2 to 1.4 mm leads to a noise level varying from 1.7 to 0.30 nm. This observation is in agreement with the expression of the standard deviation $\sigma$ of the noise amplitude derived in Ref. \cite{OC521_128577}:
\begin{equation}
\sigma = \frac{1}{8\sqrt{2}}\frac{p\Gamma}{d}\sqrt{\left(\frac{\log(N_xN_y)}{wN_\mathrm{im}}\right)},
\end{equation}
{\revision where $w$ is the full-well-capacity of the camera, $N_\mathrm{im}$ the number of averaged images, $\Gamma$ the cross-grating pitch, $p$ the dexel size, $N_xN_y$ the number of pixels of the image, and $d$ the grating-camera distance. The noise amplitude scales as the inverse of $d$.} Importantly, this $1/f^2$ non-uniformity in the spectral domain makes the noise amplitude dependent on the image size $N_x\times N_y$. The bigger the image, the larger the calculated noise amplitude. {\revision This special feature is not really a problem in CGM, because, normally, the larger the object, the thicker it is, like with living cells. However, in the case of thin and extended objects, such a low-frequency noise can be problematic. This is why the OPD values are particularly dispersed in Fig.~\ref{DMalphaMeas}(c) related to graphene, in the case of CGM, while they are not in the case of a neuron (Fig.~\ref{DMalphaMeas}(d)).}

However, the grating cannot be placed arbitrarily far, as artefacts begin to appear, as previously explained. In CGM, there exists therefore a trade-off between trueness and precision when adjusting the grating position.

CGM is inherently achromatic because it is sensitive to the shape of a wavefront, rather than the phase of light. It is commonly employed with incoherent, broad-band illumination from an LED or a bulb, sometimes in conjunction with a band-pass filter. CGM could be used with a laser illumination, but it is usually avoided because it creates the same speckle noise and fringes as the ones observed in DHM and PSI.  There are two consequences of this. First, there is no other noise origin than the shot noise. In particular, there is no speckle noise. The noise level that is observed experimentally is exactly what is observed numerically in this work \cite{OC521_128577}. Second, there is always a non-zero illumination NA. Opening the illumination NA is likely to create artefacts in QPM, like with DPM and FPM as explained further on (Figs. \ref{DPMholeSize} and \ref{FPMmaskSize}). However, it is not the case in CGM, unless large NA are used (typically above 0.6), as explained in Refs. \cite{OE17_13080,OC521_128577}. When such artefacts related to the illumination NA occur, an image post-processing can be applied to correct for the artefacts, as explained in \cite{OE17_13080}.

\subsection{Accuracy of DHM and PSI}

DHM and PSI have similar experimental implementations (Figs. \ref{QPMsetups}b,f). They represent respectively the simplest implementations of off-axis and phase-shifting methods, and exhibit perfect, artefact-free phase images for any system (Fig.~\ref{resultsTrueness}, 2nd and 4th columns). There are no inherent artefacts associated with these methods. 

These techniques also offer the possibility to adjust the relative intensities of the two arms, as a means to improve the contrast of the interferogram, a benefit that DPM does not possess for instance.\\

\begin{figure}
	\centering
		\includegraphics[scale=0.9]{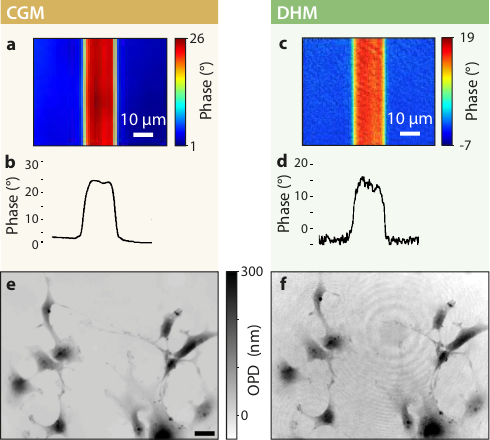}
	\caption{{\bfseries Comparison between DHM and CGM.} (a) Phase image of a waveguide acquired using CGM, along with (b) an horizontal crosscut. (c) Phase image of a waveguide acquired using DHM, along with (d) a horizontal crosscut. Reprinted with permission from ref \cite{OE26_17498}. Copyright 2018 Optical Society of America. (e) OPD image of fixed COS-7 cells obtained using CGM ($40\times$, 1.3 NA) and (f) OPD image of the same cells obtained using DHM ($20\times$, 0.7 NA). The color bar and scale bar are common to both figures. Reprinted with permission from Ref. \cite{BOE10_2768}. Copyright 2019 Optical Society of America.}
	\label{CGM-DHM-comparison}
\end{figure}

In the simulation of Fig.~\ref{resultsNoise}, DHM exhibits a very acceptable noise level compared with all the other QPM techniques. However, our simulations using IF-DDA aim to quantify the inherent inaccuracies and noise levels for each system, free from imperfections. In practice, when using a laser illumination to  make an image on a camera, the image is naturally degraded by a speckle noise and by fringes. We cannot render this type of noise in our simulations, because we model perfect illuminations and microscopes. Our numerical simulations for DHM and PSI are thus better than reality, in terms of signal-to-noise ratio. However, our simulations are not disconnected from reality because methods exist to substantially reduce the amplitude of speckle noise in DHM \cite{LSA7_48} and PSI \cite{BJ101_1025}, which is a very active research line in the QPM community.

Unlike shot-noise, or Flicker noise in CGM, a speckle noise cannot be eliminated by averaging images (unless the illumination varies over time). This noise is random in space, not in time, which is why it is particularly problematic.

In 2018, Bellanger {\it et al.} presented a comparative study of CGM and DHM \cite{OE26_17498}. One of the objectives of this article was to experimentally compare the noise levels of these two techniques. We reproduce in Figs.~\ref{CGM-DHM-comparison}(a)--(d) their results on the measurements of the phase profile of a waveguide. The effect of the coherent noise is visible on the DHM images. As a result, the noise level is higher for DHM compared with CGM. In Figs.~\ref{CGM-DHM-comparison}(e) and~\ref{CGM-DHM-comparison}(f), taken from Ref.~\cite{BOE10_2768}, also comparing CGM and DHM, the presence of fringes in DHM is evidenced.

\subsection{DPM accuracy: influence of the zero-order pinhole}\label{DPMaccuracy}

\begin{figure*}
	\centering
		\includegraphics[scale=0.9]{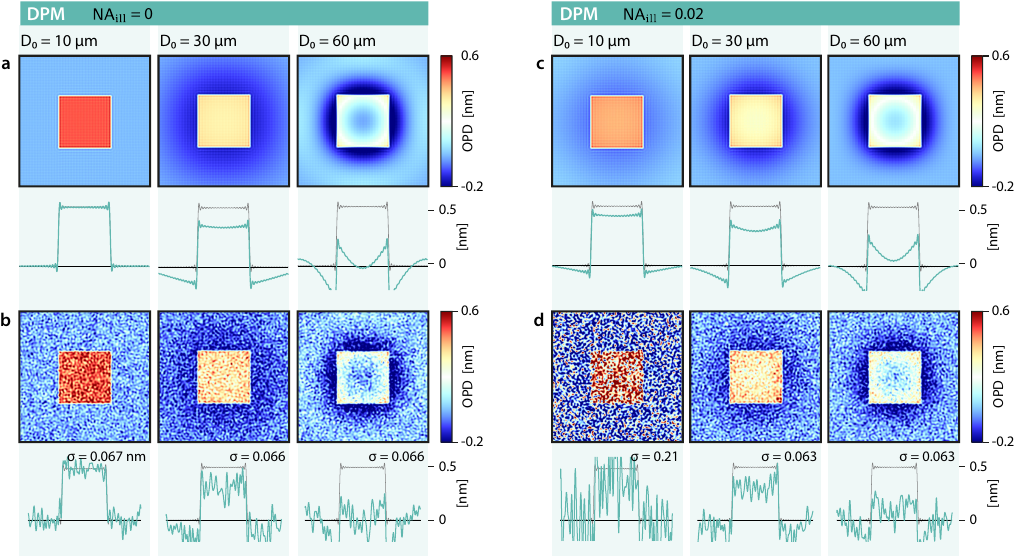}
	\caption{{\bfseries Influence of the size of the 0-order pinhole in diffraction phase microscopy (DPM) on trueness and precision}. IF-DDA simulations of OPD images acquired with DPM, without (a,c) and with (b,d) noise, along with crosscuts. While (a,b) correspond to a plane wave illumination of the sample, (c,d) corresponds to an illumination with a NA of 0.02. Each row corresponds to a different pinhole size, namely 10, 30 and 60 \textmu m.}
	\label{DPMholeSize}
\end{figure*}

DPM operates on the same principle as DHM. It is an off-axis technique that generates phase-related information through interference with a reference plane wave. However, the implementation is distinctly different, as it is a common-path method. In DHM, the reference beam follows a separate path (Fig.~\ref{QPMsetups}(c)). In DPM, the reference beam is created from a replica of the image itself, in which the lowest spatial frequency is filtered by a pinhole in the Fourier space, as a means to get a quasi plane wave (Fig.~\ref{QPMsetups}(d)). The diameter $D_0$ of the pinhole in the DPM mask is crucial. It has to be as small as possible, ideally a point, to efficiently filter all the non-zero spatial frequencies of the corresponding light profile at the image plane, to obtain a plane wave acting as the reference beam, like in DHM. When the pinhole is too large, some non-zero, low spatial frequencies are transmitted, and all these low frequencies are subtracted to the final image, leading to an area of inverted contrast around the object of interest. This effect is called the \emph{halo effect} \cite{OE22_5133,OL39_5511,OE24_11683}. The halo effect in DPM can be observed in the literature (see an example in Fig.~\ref{neurons}(d)).  Such an artefact is also observed in DPC, FPM and SLIM (Fig.~\ref{neurons}(e)) but the origin is different, as explained in the next sections. Referring to it as a \emph{halo} underestimates the issue because it implies that the problem only occurs outside the object of interest. However, each time a halo effect appears, it implies a reduction of the OPD within the object as well, sometimes called the \emph{shade-off effect} \cite{OE22_5133}, affecting the precision of any quantitative measurements, for instance of the dry mass of the cell. In Figs.~\ref{neurons}(e) and \ref{neurons}(f), the more problematic shade-off effect is even more obvious than the halo effect.

Common DPM pinhole sizes reported in the literature are 10 \textmu m \cite{ACSS5_3281,JB12_e201800291}, sometimes 15 \textmu m \cite{OLT120_105681}). Some articles even report pinhole sizes of 100 \textmu m, 150 \textmu m \cite{PNAS117_10278}, or even 200 \textmu m in \cite{OL37_1094,AOP6_57}. In Fig.~\ref{DPMholeSize}(a), we show numerical results of a 2D material imaged by DPM with various pinhole sizes, ranging from 10 \textmu m to 60 \textmu m, where the halo and shade-off effects are observed above $D_0=10$ \textmu m.  This effect of the pinhole size in DPM explains why some articles report abnormal OPD-flat eukaryotic cells and some others not: the introductory Fig.~\ref{neurons}(c) looks consistent, while Fig.~\ref{neurons}(d) looks problematic, although both of them were acquired using DPM.

Figure~\ref{DPMholeSize}(b) shows the same simulations with shot noise added to the interferogram. Interestingly, the size of the pinhole does not affect the amplitude of the noise, at least for this system consisting of a thin object.

The rule for the pinhole size in DPM appears to be 'the smaller, the better' and while generally sound, it does have limitations in at least two scenarios, which we elaborate on below: the observation of thick/large objects and the use of incoherent (white light) illumination.\\

When working with thick, scattering objects, or objects covering a large part of the field of view, {\it i.e.}, with an object scattering a large amount of the incoming light out from the direction of the optical axis, the zero-order spot in the Fourier plane becomes broad, and possibly larger than the pinhole size. In that case, if small enough, the pinhole effectively produces a plane wave, but of weak intensity compared with the first order. The two interfering light beams have thus different intensities, leading to poor fringe contrast and thus a higher noise level. 

Some studies reported the use of DPM with incoherent light sources, an approach called white-light DPM (wDPM) \cite{OL37_1094}. DPM can be much more easily used with an incoherent illumination, compared with DHM, because it is a common-path technique. In that case, the source is not a laser, but rather an LED or a bulb, and the numerical aperture of the illumination cannot be zero. As a consequence, the spot size in the Fourier plane (where the pinhole lies) is necessarily broader than with a laser. Regarding the pinhole size, there are two options: (i) either its size is big enough to capture this bigger 0-order spot. In that case, the contrast of the fringes will be optimized because the 0 and 1st orders have comparable intensities. However, there is a risk to capture non-zero, low spatial frequencies that could create halo/shaded-off artefacts, as observed in Fig.~\ref{DPMholeSize}(c) even at $D_0=10$ \textmu m; (ii) or the size of the mask is as small as possible, to avoid halo/shade-off artefacts. But in that case, the intensity of the 0-order is strongly reduced compared with the 1st order, reducing the contrast of the fringes and increasing the noise level on the images, as observed in Fig.~\ref{DPMholeSize}(d). In white-light DPM, there is a trade-off between accuracy and noise, exactly like in CGM with the grating-camera distance, but for a totally different reason.\\

Some procedures have been developed to correct for the halo effect \cite{BOE9_623}, but they are questionable, because artificially adding some low frequencies where the imaged objects are lying, and they are not always effective: For instance, in the article where Fig.~\ref{neurons}(d) was taken, the halo effect is said to be corrected, but the soma of the cell in the OPD image still appears empty. In 2021, deep learning was proposed as a method for correcting for the halo effect in DPM \cite{FP9_650108}.

{\revision 

\subsection{DPC accuracy}

The DPC simulations presented in Fig.~\ref{resultsTrueness} show significant deviation from theory, regardless of the imaged object. The theory profiles are naturally different from all the other techniques because DPC uses a small NA objective (0.4 compared with 1.3 for the other techniques). For small objects, the simulated profiles look super-resolved, while for large objects, there is a halo/shade-off effect. The two next subsections explain these observations.

\subsubsection{DPC accuracy: influence of the numerical apertures}
The only experimental degree of freedom in DPC is the NA of the 4-quadrant illumination.  It has been elucidated and experimentally demonstrated in Ref.~\cite{OE23_11394} that this illumination NA must align with the NA of the objective lens. Our IF-DDA simulations confirm this requirement. In Fig.~\ref{DPCalphaNA}, we show the image of the neuron with $(N_\mathrm{ill}, N_\mathrm{obj})=(0.4, 0.4)$ and $(N_\mathrm{ill}, N_\mathrm{obj})=(0.4, 0.8)$. In the latter case, the object totally disappears. This restriction necessitates the use of a low-NA objective, resulting in reduced spatial resolution in DPC compared to all the other QPM techniques. However, it is important to note that the effective NA in DPC equals $N_\mathrm{obj}+N_\mathrm{ill}$, {\it i.e.}, 0.8 here. This benefit enhances the initial spatial resolution by a factor of 2, but it remains low compared with the case of oil-immersion objectives. It is why the simulated DPC images look super-resolved in Fig.~\ref{resultsTrueness}, compared with theory.

\begin{figure}
	\centering
		\includegraphics[scale=0.9]{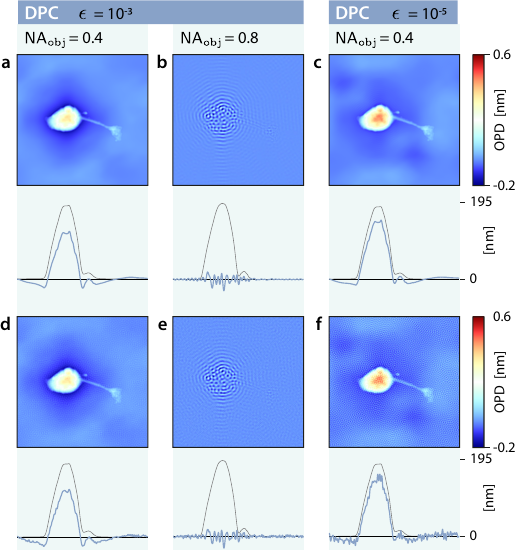}
	\caption{{\revision {\bfseries Influence of the objective NA and the Tikhonov parameter $\epsilon$ in differential phase contrast microscopy (DPC) on trueness and precision}. (a) Noise-free IF-DDA simulation of the OPD image of a neuron acquired with DPC, with an objective NA of 0.4, and $\epsilon=10^{-3}$. (b) Same as (a), with an objective NA of 0.8. (c) Same as (c) with $\epsilon=10^{-5}$. (d--f) Same as (a--c), with noise.}}
	\label{DPCalphaNA}
\end{figure}

\subsubsection{DPC accuracy: influence of Tikhonov regularization parameter $\alpha$}

The other important degree of freedom in DPC is not experimental, but numerical. The Tikhonov regularization parameter $\epsilon$ is an artificial offset in the denominator of Eq. \eqref{eq:DPCfinal}. It avoids division by small values to prevent a high noise level. The side effect of this artificial offset is a damping of the retrieved $\varphi$. Reference~\cite {OE23_11394} mentions that a good compromise is $\epsilon=10^{-3}$. It is the value we used in all the simulations presented in Figs. \ref{resultsTrueness} and \ref{resultsNoise}. In Fig.~\ref{DPCalphaNA}, we compare images of neurons obtained with $\epsilon=10^{-3}$ and $\epsilon=10^{-5}$. While the halo/shade-off effect is important for $\epsilon=10^{-3}$, it is well reduced for $\epsilon=10^{-5}$. This latter value is therefore more recommended, especially if quantitative dry mass measurements is the aim of the study. It, however, comes along with a high noise level, as observed in Figs.~\ref{DPCalphaNA}d--f, which can be discarded by average multiples images, or using a camera with a large full-well-capacity. With $\epsilon=10^{-5}$, the halo/shade-off is reduced, but some non-uniformity remains in the background.}

\subsection{FPM accuracy: influence of the size of the circular phase mask\label{FPMinaccuracy}}

\begin{figure*}
	\centering
		\includegraphics[scale=0.9]{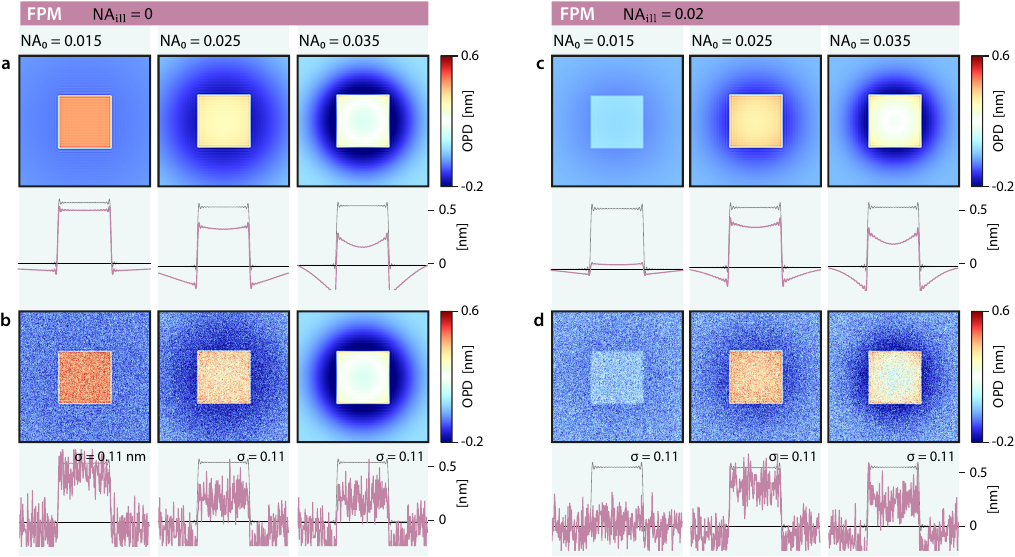}
	\caption{{\bfseries Influence of the 0-order mask diameter in Fourier phase microscopy (FPM) on trueness}. IF-DDA simulations of OPD images acquired with FPM, without (a,c) and with (b,d) noise, along with crosscuts. While (a,b) correspond to a plane wave illumination of the sample, (c,d) corresponds to an illumination with a NA of 0.02. Each row corresponds to a different numerical aperture of the phase mask $\mathsf{NA}_0$, namely 0.015, 0.025 and 0.035.}
	\label{FPMmaskSize}
\end{figure*}

Although FPM belongs to the other family of phase-shifting techniques, it is experimentally similar to DPM because it plays with the zero-order spot in the Fourier plane. However, in FPM, this spot is the only one in the Fourier space (no 1st order diffraction spots created by a grating), and it is captured by a phase mask, the dimension of which is expressed here in numerical aperture $\mathsf{NA}_0$. This similarity makes FPM suffer from similar problems as DPM, also leading to a halo/shade-off effect for some specific values of the free parameter, here $\mathsf{NA}_0$.

In FPM, the phase shifts (of 0, $\pi/2$, $\pi$,  and $3\pi/2$) are supposed to be applied to the unscattered field $E_0$, ideally only present at $\mathsf{NA}=0$ (point-like spot in the Fourier space). If $\mathsf{NA}_0$ is too large, {\it i.e.}, if the size of the phase mask created by the SLM in the Fourier space is too large, the phase shifts are not only applied to $E_0$ but also to a part of the scattered field $\mathrm{NA}_\mathrm{s}$, the low spatial-frequency part, artificially incorporating it into $E_0$, leading to a halo/shade-off effect. This issue is shown in Fig.~\ref{FPMmaskSize}(a), where such artefacts appear when the mask size is increased. This effect is all the more important for objects with low spatial frequencies, {\it i.e.}, large objects, as observed for the 2D material or the cell in Fig.~\ref{resultsTrueness}.

Just like in DPM, one could conclude that making $\mathrm{NA}_0$ as small as possible is the golden rule. However, it is not always recommended. Indeed, unlike with DPM, there is a lower limit to the free parameter $\mathrm{NA}_0$ in FPM. $\mathrm{NA}_0$ should never be smaller than the illumination NA, $\mathrm{NA}_\mathrm{ill}$. Otherwise, the phase mask is not capturing the whole $E_0$ field.  If $\mathrm{NA}_0$ is too small, such that $\mathrm{NA}_0<\mathrm{NA}_\mathrm{ill}$, then part of the incoming field $E_0$ is not phase shifted, which creates a halo/shade-off artefact. This problem occurs in particular when using incoherent (called white-light) illumination. FPM was demonstrated using white light, a modality called wFPM \cite{BOE4_1434}. In that case, the zero order is naturally not a point in the Fourier space. This problem is observed in Figure \ref{FPMmaskSize}c, with $\mathrm{NA}_\mathrm{ill}$ was set to 0.02, which is an extremely small value. For $\mathrm{NA}_0=0.015$, just below $\mathrm{NA}_\mathrm{ill}$, the signal is completely lost. This makes a strong difference with DPM, where reducing the crop never yields artefacts, and just increases the noise level. Here, with FPM, artefacts occur systematically when the phase mask is too large, but also when it is too small. And sometimes, especially with wFPM, no phase mask dimension leads to quantitative measurements. In Ref. \cite{BOE4_1434}, where wFPM was introduced, we can see that the imaged red blood cells suffer for the shade-off effect, because the central part of the RBCs feature a zero OPD value. RBCs are supposed to exhibit a lower OPD value in the center, but not a zero value. Just like wDPM, wFPM is not recommended when the quantitativeness of the measurements is important. 

Regarding the noise level, changing the phase mask size has no effect on the noise level, in any case, because it does not modify the light intensity reaching the camera sensor (Figs.~\ref{FPMmaskSize}(b) and~\ref{FPMmaskSize}(d)). This observation is much different from what happens in DPM, where a reduction of the hole size in the Fourier plane decreases the fringe contrast and thus increases the noise level (see Fig.~\ref{DPMholeSize}).

\subsection{Inherent inaccuracies of SLIM\label{SLIMinaccuracies}}

In general, in our simulations, what is assumed to be the theoretical OPD profile ({\it i.e.}, the ground truth) is the OPD profile under plane wave, normal illumination. The first problem with SLIM is that the illumination is not a plane wave, and not normal. SLIM involves an annular illumination. The question of what the definition of the true OPD should be is thus raised. In the IF-DDA simulations related to SLIM, the annular illumination is composed a 12 tilted plane waves regularly distributed along a circle in the Fourier plane at an NA of 0.25. We considered the theoretical OPD to be the average of the 12 OPD images calculated using IF-DDA, corresponding to 12 titled plane wave illuminations. This average OPD slightly differs from the OPD obtained under normal incidence (see the solid 'theory' grey lines in Fig.~\ref{resultsTrueness} for the nanoparticle simulation, compared with the other theoretical profiles on the same row). The numerical simulations show that SLIM tends indeed to measure this average OPD (Fig.~\ref{resultsTrueness}). However, this quantity can differ from what would be measured using any QPM based on a plane wave illumination.

The second problem with SLIM is the halo/shade-off effect \cite{SR7_44034}. The origin of the effect is exactly the same as in FPM (see previous section \ref{FPMinaccuracy}). SLIM encounters difficulties as soon as objects cover a large part of the field of view, because large objects feature low spatial frequencies that tend to overlap with the phase mask. The problem with SLIM is even more important compared with FPM, because in SLIM, one cannot play with the size of the mask. First, it has an annular shape, and then its dimensions has to match the dimensions of the annular phase mask of the phase-contrast objective lens. And second, this objective's annular phase mask is far from being infinitely thin. While halo artefacts can be avoided in FPM using a laser illumination and small enough a phase mask, SLIM is thus bound to feature important halo effects for large objects. Figure~\ref{resultsTrueness} illustrates this problem, where artefacts are visible for large objects such as the 2D material and the cell. Note that we used a width of 0.02NA for the annular mask, which is much better than the specification of phase contrast objectives (rather on the order of 0.1). Nevertheless, artefacts are still observed. It also explains the artefact observed in the experimental measurement of Fig.~\ref{neurons}(e). Unlike FPM, where the size of the 0-order phase mask can be reduced in size, there is no degree of freedom in SLIM, and one has to live with this artefact. SLIM looks more suited to characterise small objects, like bacteria or nanoparticles (see Fig.~\ref{resultsTrueness}).

Numerical methods have been developed to correct for the halo effect \cite{SR7_44034,BOE9_623}. However, they consist of adding a blurry image to the image, to artificially add some low frequencies and visually improves the rendering of the image. However, the claim that this trick recovers the true OPD image is not obvious, and this method tends to correct only the halo effect and not the shade-off effect.

\subsection{Effect of the defocus parameter in TIE\label{TIEinaccuracy}}
\begin{figure*}[!t]
	\centering
		\includegraphics[scale=0.9]{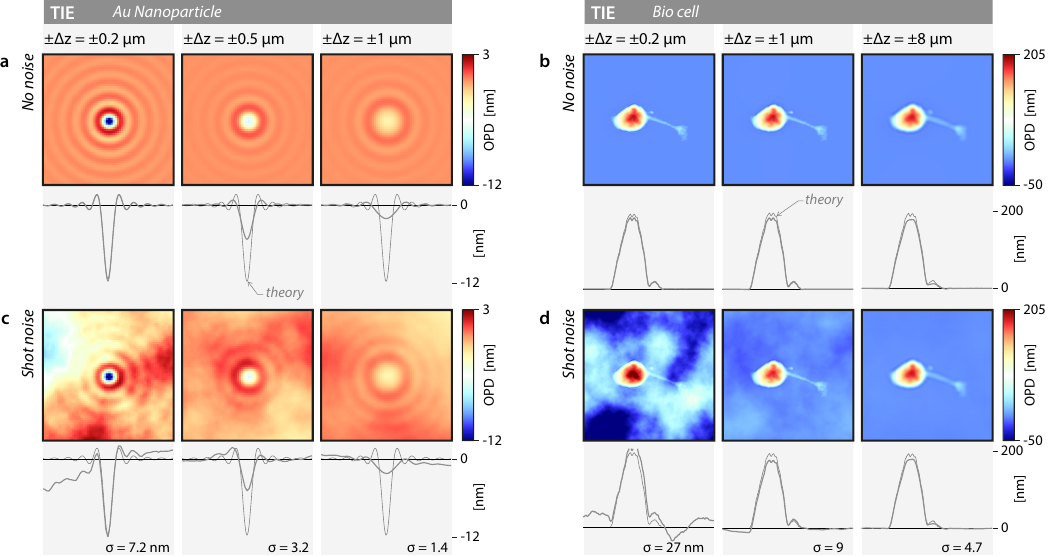}
	\caption{{\bfseries Influence of the defocus variation on the image reconstruction accuracy in transport-of-intensity equation microscopy (TIE)}. (a) Noise-free images of a nanoparticle imaged by TIE as a function of the defocus $\Delta z$, along with horizontal cross-cuts. (b) Same as (a) for a biological cell. (c,d) Same as (a,b) with shot noise.}
	\label{TIEzShift}
\end{figure*}
In TIE, three intensity images need to be acquired at various focuses ($-\Delta z$, 0 and $\Delta z$) to reconstruct an OPD image \cite{OLE135_106187}. Figure \ref{TIEzShift}(a) displays OPD images generated by TIE on a nanoparticle. In general, the smaller $\Delta z$ the better. At distances larger than 0.5 \textmu m, the OPD spreads and becomes underestimated. For larger objects, like the neuron in Fig.~\ref{TIEzShift}(b), larger defocus parameters can be used. The only issue is a loss of spatial resolution \cite{OLE135_106187}, which is evidenced by closely looking at the neurite in Fig.~\ref{TIEzShift}(b). Unlike FPM or DPM, there is no creation of artefacts with TIE  when increasing the free parameter. It just yields a loss in spatial resolution.

Figure \ref{resultsNoise} evidences a much higher degree of noise for TIE compared with the other QPMs. The reason is certainly that, unlike all the other QPMs, TIE is not an interferometry technique, in the sense that the phase map does not arise experimentally from an interference mechanism. TIE images simply arise from the subtraction of quasi-identical intensity images, $I_\mathrm{u}$ and $I_\mathrm{d}$ (Eq. \eqref{deltatzI}), making the resulting image very sensitive to shot noise from the beginning of the algorithm. Moreover, TIE primarily measures the Laplacian (2nd derivative) of the wavefront (Eq. \eqref{eq:TIE}), while CGM measures the gradient (1st derivative). Thus, two integration steps are required in TIE, which leads to a high amount of low-frequency noise ($\sim 1/f^4$), much more than CGM ($\sim 1/f^2$). As a consequence, this technique is not particularly suited {\it a priori} for imaging small or thin objects, such as nanoparticles or 2D materials. However,  the technique performs nicely for eukaryotic cells. Figures~\ref{TIEzShift}(c) and~\ref{TIEzShift}(d) show images of a nanoparticle and a living cell. To faithfully image a nanoparticle, the defocus has to remain below $\pm0.2$ $\mu$m, making the image very noisy and not really exploitable. However, for large and thick objects, such as eukaryotic cells, the defocus can be much larger without creating artefacts in the images. The drawback is rather a loss of spatial resolution on the images. This is a strong difference with other QPMs such as FPM or CGM, where the setting of the parameter comes along with a trade-off between trueness and precision.

\section{Other comparison parameters\label{QPMparameters}}
\subsection{Instabilities}
A problem that is not represented in our numerical simulations is the instabilities, which are defects that are not inherent to the technique, but coming from experimental imperfections (air flow on the setup, temperature variations, vibrations). Sometimes, they can be the main source of measurement inaccuracies. This is important because imperfection can substantially contribute to the image quality for some techniques, especially in DHM. Because there is a spatially separated arm, differences can happen between the two arms, and because of the sensitivity of interferometry measurements, even weak and unavoidable perturbations such as air flow, temperature variations or vibrations can substantially affect measurements using DHM, and even more using PSI, which is not single shot.

This problem explains the emergence of common-path techniques, which remain interferometric but discard the presence of a reference arm, such as DPM or CGM. These techniques are much less sensitive to external perturbations because the two interfering beams propagate along a common path. In CGM, the interference between the four diffracted beams occur over the caged millimetric distance between the grating and the camera, where there is no air flow. For CGM homemade systems using a relay lens, there may be a flow creating instabilities, but they can be easily discarded by shielding the beam path with tubes.

For techniques requiring multiple acquisitions (PSI, FPM, SLIM and TIE), the sample movements (including thermomechanical drifts) can be a major source of error. Parallelized acquisition to simultaneously measure the required multiple images has been proposed for TIE to overcome this limitation \cite{NP12_165}.

\subsection{Image noise}

We have explained that CGM and TIE feature a high degree of noise level, because they rely on integration steps that enhance low-spatial-frequency noise (Flicker noise). However, this type of noise can be arbitrarily reduced by averaging imaging. 

The other source of inaccuracies that cannot be taken into account in numerical simulations is the speckle noise, that occurs as soon as a laser illumination is used. It is the case in DHM, PSI, and in general for any QPM based on the use of two arms.

Unlike shot-noise, or Flicker noise, a speckle noise cannot be eliminated by averaging images. This noise is random in space, not in time, which is why it is particularly problematic.

{\revision Some approaches have been developed, both in DHM \cite{LSA7_48} and PSI \cite{BJ101_1025}, to reduce speckle noise, including optical and numerical methods. For instance, a noise reduction can be achieved with different positions of the camera or of the object, by using different longitudinal laser modes, or by using a tunable light source \cite{LSA7_48}. Also, the use of a partially coherent source is possible in DHM, as long as some caution is used \cite{OL36_2465}: the separate paths of the object and reference arms have to be carefully matched to obtain contrasted interference. Moreover, conventional DHM, using a tilted mirror like in Fig.~\ref{QPMsetups}(c), would fail in producing a contrasted interferogram throughout the full field of view, because the optical path difference cannot be uniform. The trick consists in using a grating instead of a tilted mirror. This second constraint is lifted when using PSI \cite{BJ101_1025}, but PSI does not require a tilt.}

\subsection{Speed}
Some QPMs require the acquisition of several images to reconstruct a single OPD image. It is the case of phase-shifting techniques (PSI, FPM, SLIM), which acquire 4 images, DPC which requires at least 4 images as well, and TIE, which requires 3 images. Note that more than 4 images can be acquired with DPC, to avoid possible 4-fold symmetries on the phase images \cite{OE23_11394}. Also, more than 3 images in TIE can be acquired to improve the signal-to-noise ratio. And 3 images could be sufficient in phase-shifting techniques in theory, but 4 is highly recommended to gain in signal-to-noise ratio.  Multiple acquisitions make the study of dynamical processes more complicated if they are fast. The speed of all the other QPM techniques (CGM, DPM, DHM) just equals the camera frame rate, and are considered fast for this reason.

{\revision Note that a reduced speed, due to the acquisition of multiple images, is usually not a strong drawback when investigating physical samples or even cells in culture, where the dynamical processes are usually slow.}

\subsection{Simplicity/compactness}
DHM or PSI require the full modification of a microscope, both in illumination and detection. They do not qualify as compact.

However, CGM just consists of a simple camera, to be implemented in the port of a standard wide-field microscope. It is the most compact technique one can think of, because this benefit does not exist with any other QPMs.

TIE also consists of using a simple microscope the way it is. No additional bulky system is required. It is thus particularly simple and cost effective.

DPC requires the implementation of a LED array in place of the illumination.

DPM, SLIM and FPM systems are more bulky, but they also consist of add-on modules to be adapted on the port of optical microscopes.

\subsection{Chromaticity}
The ability to vary the illumination wavelength, or the ability to use broadband illumination are valuable assets in QPM, but not all the QPM techniques have these abilities.

As soon as an SLM is used, {\it i.e.} with FPM or SLIM, a specific wavelength (or tight wavelength range) has to be used, matching the calibration of the SLM. FPM and SLIM are thus not likely to easily conduct wavelength-dependent studies. However, SLIM does not require a laser illumination. DPM and FPM can also use incoherent light sources, modalities called wDPM \cite{OL37_1094} and wFPM \cite{BOE4_1434}, but at the cost of a reduced accuracy (see Sects. \ref{DPMaccuracy} and \ref{FPMinaccuracy}).

Techniques based on laser illumination, such as DHM or PSI, are also not likely to conduct wavelength dependent studies because wavelength-adjustable lasers in the visible range and much less common and more expensive than monochromators. Monochromators and wavelength-dependent studies can however be straightforwardly conducted using CGM or TIE.

\subsection{Absolute/relative measurements}
There exist two QPM families, those actually measuring phase, and those rather imaging wavefronts (see first line of the table in Fig.~\ref{QPMcategories}). The wavefront imaging techniques are CGM and TIE. Because they firstly image wavefront gradients that are eventually integrated, an additive constant appears, making the measurements not absolute. In other words, the OPD map is obtained to within a constant. Consequently, CGM and TIE are not sensitive to global variations of the phase over the field of view, also called a piston. Indeed, in TIE, a global phase variation does not modify any of the acquired intensity images $I_\mathrm{u}$, $I_\mathrm{d}$ and $I_0$, precisely because they are intensity images. In CGM, a global phase variation does not modify the spot-array pattern of the interferogram. Albeit a phase imaging technique, DPC also provides relative measurements because it primarily measures phase gradients. Indeed, DPC is not interferometric and any piston does not modify the set of regular intensity images acquired with DPC. However, in DHM for instance, progressively varying the global phase over the field of view results in a progressive fringe shifting in the interferogram. Thus, CGM and TIE (wavefront imaging techniques) provide wavefront profiles to within a constant, while phase imaging techniques are capable of measuring absolute phase variations, in theory.

However, getting the phase/wavefront to within a constant is not a limitation in most applications, especially in cell biology. For instance, dry masses can be still easily measured in CGM considering that the outer boundary of the object in the OPD image is zero \cite{JBO20_126009,BOE13_6550,BJ122_1}.

\subsection{Polarisation}

Some samples can be birefringent, meaning that their refractive index depends on the polarisation of the incident light. It is the case of collagen fibers, or anisotropic nanoparticles for instance. Birefringence characterisation is possible using techniques such as CGM \cite{OE23_16383,OC422_17} or DHM \cite{BOE13_805} by rotating the polarisation of the incident illumination, or by implementing a polarizer in detection. However, any QPM based on the use of an SLM ({\it i.e.} SLIM and FPM) cannot study birefringence because an SLM requires a linearly polarised light beam along a given direction. 

\subsection{Spatial resolution and image definition}

{\revision CGM, DHM and DPM belong to the same family of off-axis QPMs. They all rely on the acquisition of an interferogram, characterized by fringes, from which the OPD image can be retrieved from a single image, at the cost of a reduction of the number of pixels of the image, compared to the camera raw image. One usually has an area of $3\times3$ pixels of the raw image (interferogram) that corresponds to 1 OPD pixel \cite{ACSP10_322,OC521_128577}.  This reduction of image definition does not necessarily lead to a reduction of the image resolution. If the microscope oversamples the object, then the diffraction limit can still be reached \cite{BJ106_1588,AO53_2058,S_2304564}.

On the contrary, DPC, PSI, FPM, SLIM, TIE do not gain information by sacrifying some image definition, but by acquiring multiple images. Thus, with these techniques, the processed OPD and intensity images have the same number of pixels as the raw camera image.

DPC is special in the sense that it requires a low-NA objective, typically with a NA value of 0.4. Since the illumination NA has to match the objective NA, the effective NA in DPC is twice as large, {\it i.e.} 0.8. However, it is still smaller than what the other QPMs can do. The spatial resolution in DPC is thus usually poorer than with other QPMs. \\}

\begin{figure}
	\centering
		\includegraphics[scale=0.9]{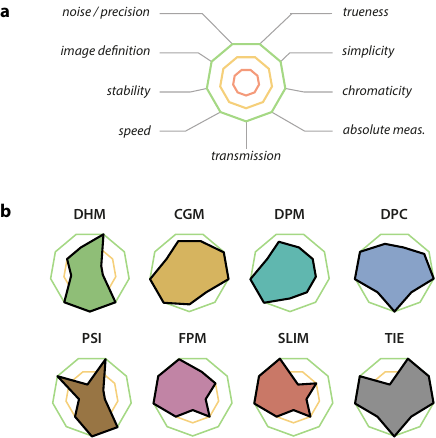}
	\caption{{\bfseries Comparison of QPM features.} (a) Designation of the features of the radar charts. (b) Radar chart of the 8 QPM techniques.}
	\label{radarChart}
\end{figure}

\begin{figure*}
	\centering
		\includegraphics[scale=0.9]{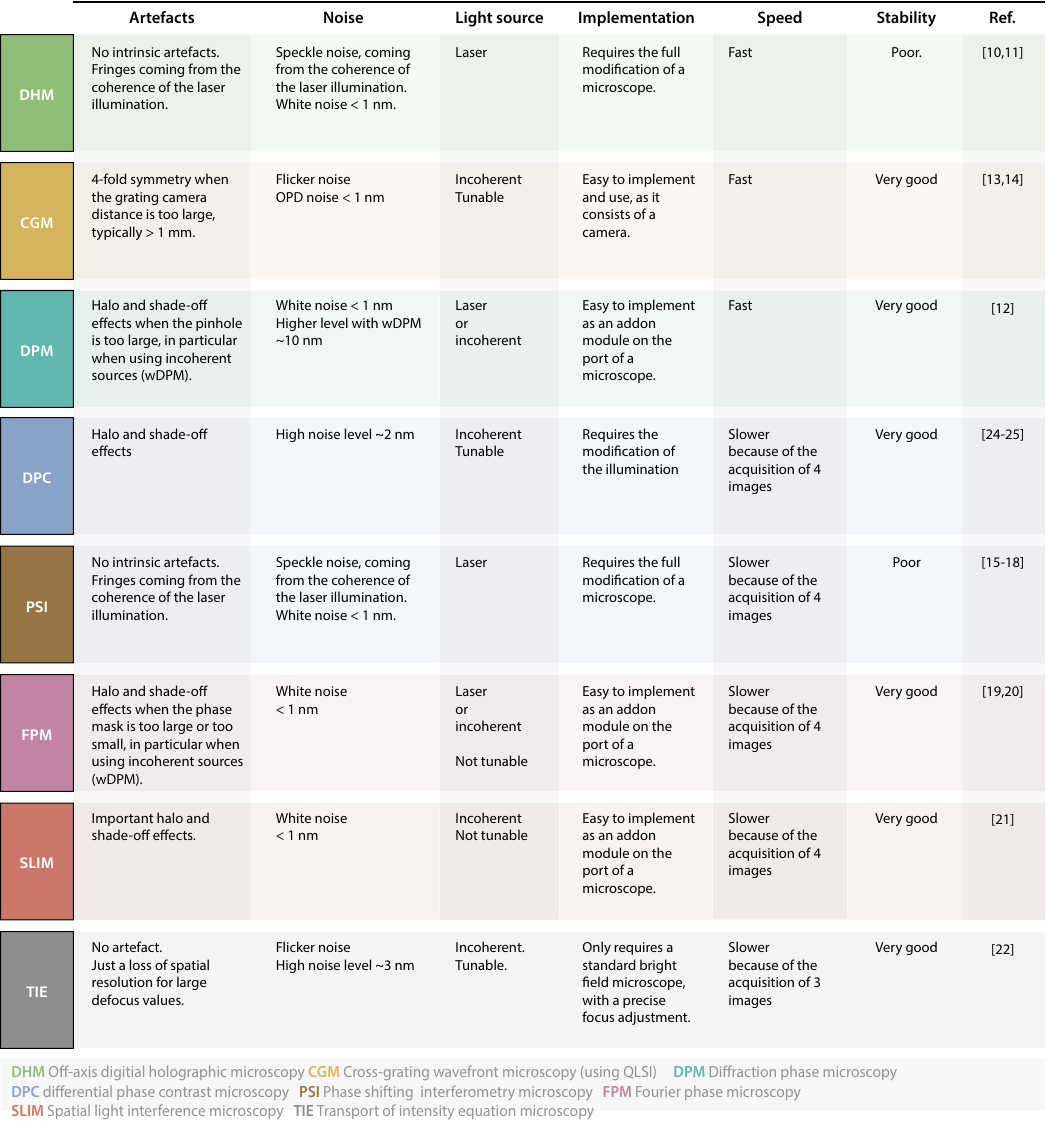}
	\caption{{\bfseries Summary of the main QPM benefits and drawbacks.}}
	\label{QPMtable}
\end{figure*}

\section{Summary}

The aim of the article is twofold, (i) present and share a numerical toolbox tailored to modeling any QPM, and (ii) draw comparisons between important QPM techniques, as a guide to help the reader find their way in the field of QPM and select the most appropriate technique for their particular application.

Figure \ref{radarChart} visually summarises the results in the form of radar charts, as a means to better compare the features of all the 8 QPM techniques. 9 parameters are included, namely the precision and the trueness (following the results described in Sect. \ref{results}), and the simplicity, chromaticity, absolute nature of the measurements, transmission, speed, stability and image definition (following the results described in Sect. \ref{QPMparameters}). Evaluations are made on 3 grades, following the results and the discussion presented in this article. Noteworthily, some of these seven patterns are geometrically nested within others, indicating that certain techniques lack interest compared to others.

{\revision It is worth comparing DPM and CGM, which are the two grating-based techniques. The basic version of DPM does not offer clear benefits compared to CGM. The noise of DPM is just better in theory, but in practice it suffers from speckle noise. However, DPM presents interest when accessing Fourier space is necessary, as in this implementation of single-shot, two-color phase measurements \cite{M136_35}.

Also, the SLIM pattern is encompassed by the FPM pattern, the two PSI methods based on the use of an SLM, meaning that SLIM does not offer significant benefits compared to FPM. {\revision It is also not obvious what the interest of SLIM is compared with phase-contrast microscopy, especially for eukaryotic cells. The official interest of SLIM is to turn phase-contrast images into quantitative OPD images. But as demonstrated in this study, images of eukaryotic cells are hardly quantitative in SLIM.}

In general, the appropriateness of a method compared to another also depends on the imaged object. For large objects such as eukaryotic cells (neurons, red blood cells, cancer cells, etc), we do not recommend DPC, FPM and SLIM if quantitative dry mass measurements are the aim, because of halo/shade-off effects that tend to underestimate the dry mass.

As a means to summarize the results in a less visual but more detailed manner, Fig.~\ref{QPMtable} presents a table that includes brief comments on all the features of all the QPM techniques described in this article.

Finally, note that our numerical tool could also be used to compare 3D tomographic reconstructions provided by various QPM techniques. Indeed, IF-DDA can calculate the total field in any defocused plane with respect to the microscope focal plane, and for any illumination angle. With such data, 3D reconstructions can be easily studied.}

\bmhead{Funding}
This project received funding from the European Research Council (ERC) under the European Union’s HORIZON Research and Innovation Programme (proposal number no. 101101026, project MultiPhase).

\bmhead{Data availability}
All data required to reproduce the results can be obtained from the corresponding author upon a reasonable request.

\bmhead{Code availability}
The Fortran, C++ and Matlab codes used in this study are accessible on the internet:
\begin{itemize}
\item The IF-DDA toolbox to compute rigorously the image of an object through a microscope can be downloaded from gitlab~\cite{IFDDAgitlab} or from the webpage: 

\burl{https://www.fresnel.fr/spip/spip.php?article2735&lang=fr}

Two codes are available; one where objects are in free space and the other where objects may be in the presence of a multilayer. This is the one used in this article to take the substrate into account.

\item The Matlab add-on to the IF-DDA toolbox to implement the QPM configuration is accessible from github~\cite{GitHub-CGM-QPM}.
\end{itemize}

\bmhead{Supplementary Information} The online version contains supplementary material available at https://xxxxxxxxxxxxx.

\bibliographystyle{unsrt}

\end{document}